\pdfoutput=1

\documentclass[acmtog]{acmart}

\usepackage{booktabs} 
\usepackage[normalem]{ulem}

\author{Stavros Diolatzis}
\affiliation{
	\institution{Inria \& Université Côte d'Azur}
	\department{GraphDeco}
	\city{Sophia Antipolis}
	\postcode{06902}
	\country{France}}
\email{stavros.diolatzis@inria.fr}
\author{Julien Philip}
\affiliation{
	\institution{Inria \& Université Côte d'Azur}
	\department{GraphDeco}
	\city{Sophia Antipolis}
	\postcode{06902}
	\country{France}}
\affiliation{
	\institution{Adobe Research}
	\city{London}
	\country{UK}}
\email{juphilip@adobe.com}
\author{George Drettakis}
\affiliation{
	\institution{Inria \& Université Côte d'Azur}
	\department{GraphDeco}
	\city{Sophia Antipolis}
	\postcode{06902}
	\country{France}}
\email{george.drettakis@inria.fr}

\usepackage[ruled]{algorithm2e} 
\usepackage{overpic}

\SetAlFnt{\small}
\SetAlCapFnt{\small}
\SetAlCapNameFnt{\small}
\SetAlCapHSkip{0pt}

\acmJournal{TOG}

\setcopyright{licensedothergov}
\acmYear{2022} \acmVolume{41} \acmNumber{1} \acmArticle{1} \acmMonth{1} \acmPrice{15.00}\acmDOI{10.1145/3522735}

\definecolor{copper}{rgb}{0.72, 0.45, 0.2}
\newcommand{\NEW}  [1] {}
\newcommand{\NEWTEXT}  [1] {#1}
\newcommand{\REM}  [1] {\ignorespaces}

\usepackage{array}
\newcommand{\PreserveBackslash}[1]{\let\temp=\\#1\let\\=\temp}
\newcolumntype{C}[1]{>{\PreserveBackslash\centering}p{#1}}
\newcolumntype{R}[1]{>{\PreserveBackslash\raggedleft}p{#1}}
\newcolumntype{L}[1]{>{\PreserveBackslash\raggedright}p{#1}}

\def\NNtype{PixelGenerator~}
\def\TRAINTIMERANGE{5-18 hours}
\def\MAXTRAIN{18 hours}
\def\OURFPS{4-6 fps}

\def\x{x}
\def\LN{Loss_{\mathrm{new}}}
\def\LE{Loss_{\mathrm{exist}}}
\def\D{$\mathcal{D}$~}
\def\Dn{$\mathcal{D}$}
\def\w{\omega}
\def\prod{\mathcal{P}}
\def\bsdf{\rho}

\let\oldput\put
\def\put(#1,#2)#3{%
	\oldput(#1,#2){\sffamily #3}%
}

\begin{document}
	\title{Active Exploration for Neural Global Illumination of Variable Scenes}
	
\begin{abstract}

Neural rendering algorithms introduce a fundamentally new approach for photorealistic rendering, typically by learning a neural representation of illumination on large numbers of ground truth images. When training for a given \emph{variable} scene, i.e., changing objects, materials, lights and viewpoint, the space \D of possible training data instances quickly becomes unmanageable as the dimensions of variable parameters increase. We introduce a novel \emph{Active Exploration} method using Markov Chain Monte Carlo, which \emph{explores} \D, generating samples (i.e., ground truth renderings) that best help training and interleaves training and on-the-fly sample data generation. We introduce a self-tuning sample reuse strategy to minimize the expensive step of rendering training samples. We apply our approach on a neural generator that learns to render novel scene instances given an explicit parameterization of the scene configuration. Our results show that Active Exploration trains our network much more efficiently than uniformly sampling, and together with our resolution enhancement approach, achieves better quality than uniform sampling at convergence. Our method allows interactive rendering of hard light transport paths (e.g., complex caustics) -- that require very high samples counts to be captured -- and provides dynamic scene navigation and manipulation, after training for \TRAINTIMERANGE~depending on required quality and variations.

\end{abstract}


\begin{teaserfigure}
	\centering
\vspace{-3mm}
	\begin{overpic}[width=\linewidth]{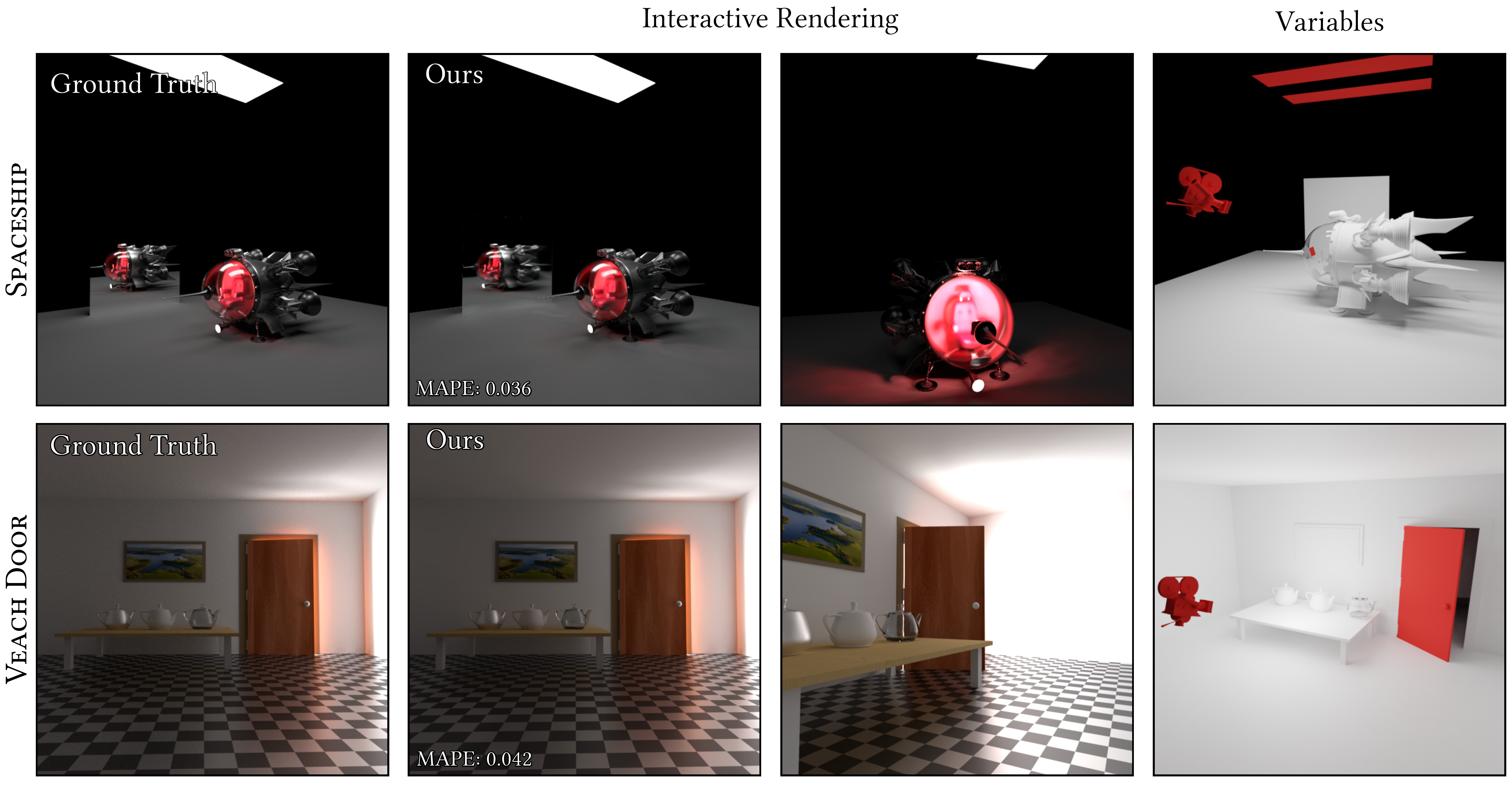}
	\end{overpic}
	\caption{\label{fig:teaser} 
	\NEW{Updated GTs.} We introduce a neural rendering method that allows interactive navigation in a scene with dynamically changing properties, i.e., viewpoint, materials and geometry position and full global illumination effects. With our \emph{Active Exploration} we can train a neural network \NEWTEXT{efficiently} to learn global illumination for all the configurations of these variable properties, allowing interactive rendering at runtime.
	Left to right: ground truth path traced images; our prototype interactive neural renderer, running at \OURFPS~ \NEWTEXT{with a variation of each scene and the variable parts of the scene depicted in red; each variable property (light intensity, camera position, object rotation, etc.)} is controlled by an interactive slider (please see video). }
\end{teaserfigure}
	
\maketitle
	
\section{Introduction}

Neural rendering is an exciting emerging alternative to traditional physically-based rendering; Initial attempts \cite{ren2013} were restricted to indirect light with static geometry, while more recent methods  \cite{eslami2018neural,granskog2020compositional} are trained on large numbers of rendered images for variable scenes, i.e., with changing objects, materials, lights and viewpoint. 
Other recent solutions learn encodings of appearance or lighting~\cite{nalbach2017deepshading,baatz21nerftex,zhu21luminaires}, again training on rendered images. Uniformly sampling the space of these path-traced images is expensive; in the case of a high-dimensional space \D containing all the possible configurations of a variable scene, it quickly becomes unmanageable. To address this limitation, we introduce an \emph{Active Exploration} strategy, that guides sampling to parts of the space \D that are most challenging for neural rendering.

We demonstrate the efficiency of our Active Exploration approach on a neural renderer, by training a generator network that can interactively render global illumination with dynamic modifications (moving viewpoint, lights, objects etc.).\NEWTEXT{ Training time varies from} minutes to hours, depending on the scene variability, complexity and available hardware. To represent a variable scene (see Fig.~\ref{fig:teaser}), we use an explicit representation with a vector $v$ of variable parameters that precisely define an instance of the possible scene configurations in \D, enabling fine control of each parameter and interactive rendering. 

We \emph{interleave} training with on-the-fly generation of the data it needs.
Uniformly sampling the space of parameters to generate the data for training does not allow the network to achieve satisfactory visual quality, especially when increasing the dimensions of \Dn, because in many cases light transport has hard, localized effects that have low probability of being observed. 

Our Active Exploration method finds samples best suited for training but also locally explores these regions of \D which is crucial in our context especially for enabling high resolution training (\ref{sec:resolution}), compared to Active Learning (see Sec.~\ref{sec:related}). For this we use a Markov Chain Monte Carlo (MCMC) approach, with small and large steps and a custom acceptance policy.

Despite our focused Active Exploration, training data generation -- i.e., ground truth path-tracing -- is still expensive; it is thus beneficial to \emph{reuse} such rendered samples during training. For best results, we introduce a \emph{self-tuning sample reuse} strategy that optimizes the probability for training sample reuse, further reducing training time.

Active Exploration, together with sample reuse and resolution enhancement allow us to train our neural renderer network very efficiently. In contrast, uniform sampling converges to a low quality solution, while our guided exploration of the training space allows us to significantly improve visual quality, especially for reflections and hard light paths. Therefore our renderer is well suited for interactive rendering of effects such as complex caustics or specular-diffuse-specular paths, that are not handled by other real time methods.
 
In summary our contributions are:
	\begin{itemize}
	\item A novel Active Exploration approach, interleaving training with on-the-fly generation of training data, together with an adaptively increasing resolution method.
	\item A self-tuning sample reuse approach, further optimizing training time and storage.
	\item A neural renderer that allows direct control of parameters for global illumination and interactive inference, based on an explicit scene parameterization.
	\end{itemize}

We demonstrate our system that allows interactive modifications of lighting, geometry, materials and viewpoint (at \OURFPS ~in our prototype Python implementation, see Fig.~\ref{fig:teaser} and video). We will provide all source code and data on publication.

\section{Related Work}
\label{sec:related}
	
We review the most closely related work in traditional and neural rendering, and discuss some aspects of deep learning that inspired our Active Exploration and training sample reuse methods.

\subsection{Traditional Global Illumination}

Monte Carlo methods, and in particular path-tracing and variants, are now the standard method for realistic, physically-based rendering, used extensively in industry~\cite{keller2015path}. In this paper we use the modern GPU-accelerated Mitsuba 2 path tracer~\cite{nimier2019mitsuba} for generation of training data. 

Markov Chains are used extensively in Monte Carlo rendering since they converge to generating samples according to their importance by only evaluating the contribution of each path. Veach and Guibas~\shortcite{veach1997metropolis} used Metropolis Sampling to explore the space of all possible paths. Kelemen et al.~\shortcite{kelemen2002simple} later applied the same sampling in the space of random numbers that create the paths instead of the paths themselves, i.e., in Primary Sample Space. The space of random numbers that we use to create each scene instance during the data generation of neural rendering has many similarities with this approach: They are both in general high dimensional and for both importance is distributed unevenly in pockets of each space. Also related to our method is the work by Bitterli et al.~\shortcite{bitterli2019selectively} that combines a simple path tracing integrator with MCMC by using the random seeds of high variance paths as starting points for the Markov Chains.
	
Numerous methods have been developed for interactive global illumination approximations~\cite{ritschel2012state}. These methods are typically approximations either of light transport, e.g., simulating only one bounce of indirect lighting~\cite{ritschel2009approximating} or limiting light transport to a subset of possible paths~(e.g., diffuse-only \cite{nichols2009hierarchical}).
Dynamic scenes (moving objects or lights) are rarely treated in such methods, e.g., the method of Majercik et al.~\shortcite{majercik2019dynamic}, that is limited to diffuse scenes. 

One popular approach is the use of light probes~\cite{greger1998irradiance,mcguire2017real,RLPWSD20}, and in some cases 
limited dynamic behavior is possible (e.g., dynamic lights~\cite{silvennoinen2017real}).
Seyb et al.~\shortcite{seyb2020design} provides a solution for a scene with variations -- or tunable parameters --  using extensive precomputation, demonstrating the importance of this use case in production. 
For such methods, overall accuracy often depends on probe density and limitations of the probe representation and reconstruction method, and they are usually restricted in the type of light transport supported (e.g., diffuse only).
In contrast, we learn complete light transport for a variable scene, avoiding the explicit representation and reconstruction of lighting at test time and using Active Exploration to find the most useful ground truth, fully path-traced samples.

Recent hardware advances, and in particular ray-tracing hardware~\cite{Nvidia-RTX,bentyfalcor} open a new horizon for interactive global illumination, including resampling algorithms that greatly accelerate real-time rendering of very complex, dynamic direct lighting~\cite{bitterli2020spatiotemporal}. 
Nonetheless, complex global light transport paths are hard to compute on-the-fly, suggesting that precomputation-based solutions could be combined with real-time ray-tracing for future interactive Global Illumination algorithms. In the results (Sec.~\ref{sec:results}) we show examples of difficult light transport configurations that can be rendered interactively with our method.

\subsection{Neural Rendering}
\label{sec:prev-neural}

Using neural networks for rendering is a rapidly growing research topic with applications in many different contexts. The area is vast; we briefly review a few representative papers, and refer the reader to a survey~\cite{Tewari2020NeuralSTAR} for more information.

The most widespread usage of neural rendering has been in the context of \emph{inverse} problems, i.e., capturing and rendering real scenes. Even though we treat only synthetic scenes, the methodologies developed in this context can be applied in our case.  Methods vary from 3D reconstructed geometry-aware solutions for rendering~\cite{HPPFDB18,riegler2020free} or multi-plane methods~\cite{srinivasan2019pushing} to continuous volumetric representations ~\cite{mildenhall2020nerf}. SIREN~\cite{sitzmann2019siren} uses sinusoidal activation functions that greatly improve reconstruction quality.
Such view-synthesis methods are restricted to the lighting condition at capture. 
More recent methods allow modification of lighting directly~\cite{PGZED19}, and a recent method represents materials, geometry and lighting to estimate and modify lighting~\cite{srinivasan2020nerv}. X-Fields~\cite{Bemana2020xfields} also map view, time and light coordinates for small-baseline camera motion.
These methods use ideas from global illumination methods and have inspired aspects of our work, e.g., our choice of the neural network. They focus mostly on real scenes, while we work exclusively with \emph{synthetic} content.

For such synthetic content, there has been significant effort to improve sampling~\cite{muller2019neural} or post-process denoising~\cite{bako2017kernel, gharbi2019sample, chaitanya2017interactive, isik2021anf} for Monte Carlo global illumination algorithms. Denoising methods fail to reproduce complex lighting effects (caustics, specular-diffuse-specular paths) if they do not exist in the noisy input, while our method can reproduce them at interactive rates in high resolution (see Sec.~\ref{sec:comp-anf}).
\NEWTEXT{Deep learning methods have been used to improve importance sampling~\cite{Bako19,zheng2019learning}.}
In more recent work M{\"u}ller et al.~\shortcite{muller2019neural} use a neural importance sampler that is trained in an online fashion to guide the training, similar to our exploration scheme. The basic premise for the Noise2Noise approach of Lehtinen et al. ~\shortcite{lehtinen2018noise2noise} helps guide the level of image quality required for our on-the-fly path-traced training data (see Sec.~\ref{sec:arch-buff})

Our work was inspired by the use of neural networks to replace or augment rendering with global illumination. 
In pioneering work, Ren et al.~\shortcite{ren2013} learned indirect illumination with dynamic lights and roughness using neural networks, trained on a fixed set of precomputed renderings. While sharing similarities to our method, our Active Exploration method allows the treatment of more variability (e.g., moving objects) since it efficiently identifies important samples as the dimensionality of the training space increases. Our adaptive resolution approach also allows direct lighting to be learned. 

A later solution learned screen-space shading effects ~\cite{nalbach2017deepshading}. More recently, Eslami et al.~\shortcite{eslami2018neural} train an encoder decoder network on a variable scene by encoding observations into a scene representation vector, which is then used to render the scene from a novel view point.
Granskog et al.~\shortcite{granskog2020compositional} expand on this idea by using buffers to help the network and by enforcing structure on the neural scene representation. Such representations are compact compared to traditional representations, such as voxel grids and point clouds, scaling more gracefully with scene complexity and size. 
For Granskog et al.~\shortcite{granskog2020compositional} generating a novel configuration still requires rendering three observations, i.e., full path-traced images. 

The Neural Radiance Cache by Mueller et al.~\shortcite{muller2021real} uses a smaller neural network with online training, that is used to query indirect illumination for real-time path tracing. Like denoising methods, it depends on noisy real-time path tracing that often misses difficult light transport paths which our method handles well.
\REM{In work to appear, }Hadadan et al.~\shortcite{hadadan2021neural} propose to minimize the rendering equation residual using a neural network. Similar to us, they use scene information such as material parameters and normals to aid the network, but they \NEWTEXT{cannot} handle variable scenes, \NEWTEXT{without retraining}, which is the focus of our method. Additionally, during training they uniformly sample positions and incoming directions on the scene surfaces, compared to our Active Exploration.

\begin{figure*}[!h]
\begin{overpic}[width=\linewidth]{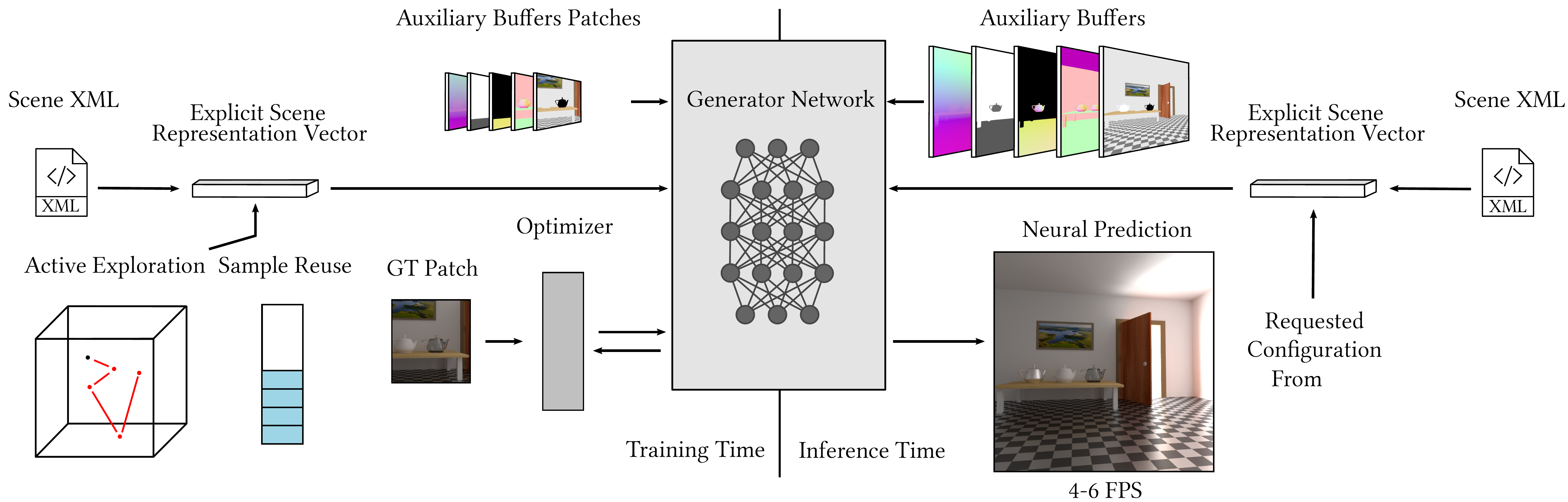}
	\put(1.5,12){$\mathcal{D}$}
	\put(11.4,20.5){$v$}
	\put(78.7,20.5){$v$}
	\put(85,7.42){$\mathcal{D}$}
\end{overpic}
\caption{\label{fig:overview}
	Overview of our approach. Left: During training we define a scene and the set of variable parameters via an \textit{xml} file resulting in a explicit scene representation vector $v$. Using Active Exploration we guide the configurations of the \emph{variable scene} towards more difficult instances that are important for the \NNtype network. Right: After \TRAINTIMERANGE ~of training -- depending on the complexity of the variable parameters and quality required -- we can \emph{interactively} request any variation of the scene with visual dynamic changes in illumination, move objects, the viewpoint and modify materials.}
\end{figure*}
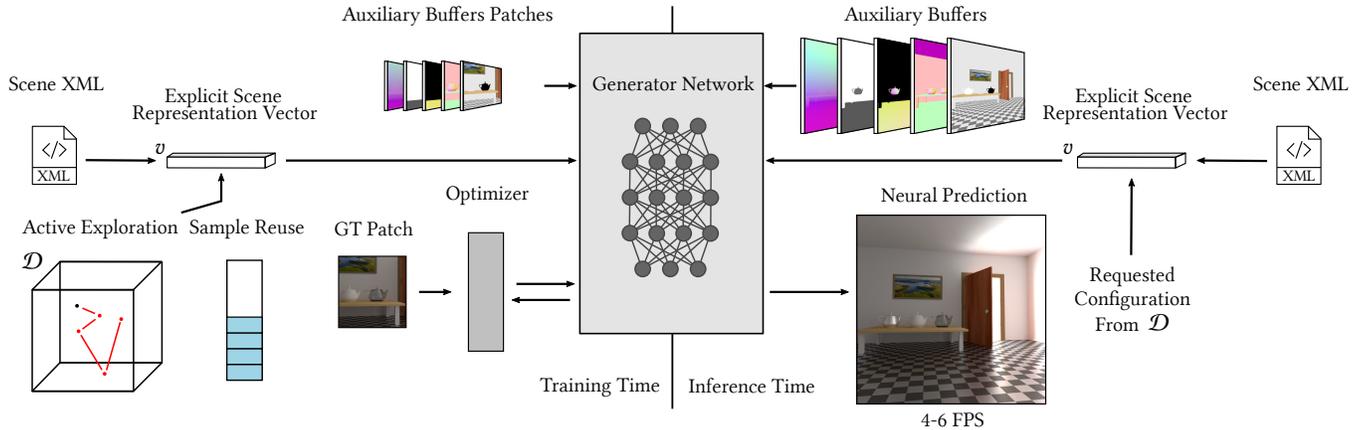
	
\subsection{Machine Learning}

Our on-the-fly data generation, Active Exploration and training sample reuse approach do not have obvious equivalents in related work to our knowledge. However, several sub-fields of machine learning explore ideas with some similarities; we review these briefly.

\textit{Active Learning.}~
Parallels can be drawn between our on-the-fly training data generation and Active Learning, where the data generation (labeling) is done procedurally to decrease cost. As reviewed by Settles~\shortcite{settles2009active} in Active Learning an algorithm chooses when a data sample needs to be labeled, i.e., to be given the ground truth. Active learning has been applied to convolutional neural networks~\cite{sener2017active} and generative adversarial networks~\cite{zhu2017generative}. Different metrics can be used to define the importance of each sample; some are related to our metrics to identify the most important samples during Active Exploration (Sec.~\ref{sec:MCMC}). Our context of working with synthetic scenes allows us to expand on Active Learning and introduce Active Exploration. We not only identify hard samples for training but we also use mutations to propose new hard samples which helps with catastrophic forgetting~\cite{kemker2018measuring} and overfitting.
	
\textit{Curriculum Learning.}~
Importance sampling methods with Stochastic Gradient descent have been developed 
under the general curriculum learning framework~\cite{bouchard2015online}. They learn the probability distribution of choosing a training sample and use it for importance sampling. 
Similarly Hazan et al. \shortcite{hazan2011beating} learn a distribution for picking training data.
In contrast to such methods, we know the exact dimensions of our data space and can sample them at will, making the task easier.

Recent work investigates issues with adaptive sampling, and the cost of using the ideal target function~\cite{stich2017safe}, and provide guarantees about the quality of sampling given limited information on the gradients. 
\NEWTEXT{There have been some techniques that use self-augmentation with synthetic rendering to overcome the lack of labelled or real-world ground truth data~\cite{kim2018inversefacenet,li2017modeling}; the goal is to match the synthetic and real distributions, which implies different design choices from our context.}

Compared to all these methods the major difference is that we have a forward problem, and thus have full knowledge of the parameters that define the space of training data and their dimensions. We can thus sample any part of this space on-the-fly.
This aspect of the space of training data makes it amenable to an MCMC exploration method, which is not the case of static, pre-captured training datasets.

\textit{Learning and MCMC.}
MCMC methods have been used in Bayesian learning from the early days of neural networks~\cite{neal1996bayesian}. More recently, Stochastic-Gradient MCMC has been proposed \cite{welling2011bayesian,zhang2019cyclical} with various applications~\cite{li2016learning}. We also use MCMC for deep learning, but in a different context: since we solve a \emph{forward} problem and can generate training samples on-the-fly, we use an MCMC approach inspired by Metropolis-Hastings to guide the sampling process.

\section{Overview}

Our goal is to significantly improve the efficiency of the training process in neural renderers that are trained on synthetic rendered data by introducing Active Exploration of the high-dimensional sample space and re-using these samples. With this scheme we are able to efficiently train our neural renderer which provides explicit control of the scene parameters and has constant rendering performance regardless of the difficulty in the underlying lighting effect.

We represent the scene variability by an explicit scene parameter vector $v$ (Fig.~\ref{fig:overview}), which
defines the space \D of all possible configurations of the scene; thus any $v~\in~$\D corresponds to an individual scene configuration.  Our goal is to train a network to take a specific $v$ and the set of corresponding G-buffer images (normals, albedo, etc.) as input, and generate full global illumination images (Inference Time in Fig.~\ref{fig:overview}).

When training the network some visually significant effects are very localized in the high dimensional space \Dn. Finding sufficiently useful samples in \D to train our network for those effects is very unlikely using uniform sampling and our limited budget. It becomes more unlikely as the dimensionality of \D grows.

Since we are solving a \emph{forward} problem, we can generate ground truth training samples on-the-fly using a fast path-tracer. A \emph{training batch} will be 16 \emph{samples} each consisting of a 32x32 patch of ground truth, path-traced image, each patch in the batch corresponding to a different configuration of the vector $v$. Each patch is sampled by a different Markov Chain and rendered in parallel. Using patches allows more efficient exploration of \Dn. In terms of pixels generated, this is equivalent to an image of resolution 128x128.

Even though our path tracer is fast, the cost of generating a training batch
is still high. We thus \emph{reuse} training samples as much as possible. We introduce a sample reuse strategy that further improves the speed of training. Training times vary from \TRAINTIMERANGE~ depending on the complexity of the variable parameters and the quality required.

Once trained, the generator network allows interactive rendering of dynamic global illumination effects (Fig.~\ref{fig:overview}, right) for the variable scene, e.g., interactively navigating in the scene, opening the door, change lighting etc. (please see video).

\section{Explicit Encoding and On-the-fly Data Generation}

We explain the explicit scene representation, the generator network and the on-the-fly data generation process, used in our neural rendering algorithm, before presenting the actual Active Exploration approach in Sec.~\ref{sec:active-exp}.

\subsection{Explicit Scene Representation}

Previous methods \cite{eslami2018neural}\cite{granskog2020compositional} use an encoder network to create a neural scene representation vector of a scene configuration (Sec.~\ref{sec:prev-neural}).

However, this representation lacks interpretability and editability. In addition, rendering a new scene configuration requires new observations (i.e., ground truth renderings) to be generated,  since the parameters of the scene representation have no explicit interpretation or ``meaning'', and thus the renderings are needed to generate the new neural scene representation vector.

We focus on variable scenes, commonly used in production~\cite{seyb2020design}. 
Given that we know explicitly which parts of a scene are variable and how much they can vary, we avoid training an encoder network to represent this variability and instead create the scene representation vector from the scene definition. As a result all fixed properties are stored in the generator and associated with a set of inexpensive G-buffers, while scene variability is compactly represented in the explicit vector. This vector contains the normalized values of the variable scene parameters for a given scene instance.

\begin{figure}[!t]
	\includegraphics[width=\linewidth]{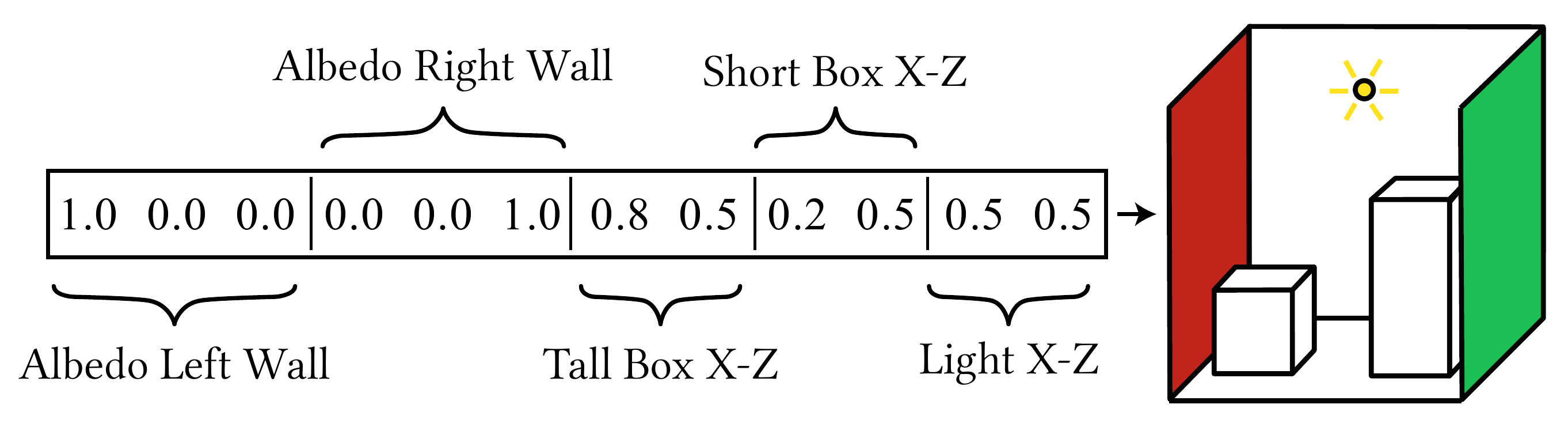}
	
	\caption{ The explicit scene representation vector $v$ that defines this instance of the variable Cornell box scene.}
	\label{fig:vector}
\end{figure}

Consider a variable Cornell box scene (Fig.~\ref{fig:vector}). Here we vary the two vertical wall albedos, positions of the boxes and light source position, and define the ranges of these parameters. 

The normalized parameter values make up the explicit scene representation vector $v$ (Fig.~ \ref{fig:vector}). The scene representation vector along with the camera position and lookat vector are repeated along the width and height dimension to be the same size as the G-buffers, and also passed to the neural network. \NEWTEXT{Since our generator operates on a per pixel basis, this repeated vector injects the global scene information to all pixels}.

\subsection{Network Architecture, Buffers and Training Data}
\label{sec:arch-buff}

We optimize a modified \NNtype architecture~\cite{granskog2020compositional,sitzmann2019srns} (a Multilayer Perceptron network with skip connections) to map the inputs for each pixel to the final pixel color value. We choose this over a convolutional neural network such as a UNet since \cite{granskog2020compositional} has demonstrated the \NNtype architecture to perform better at upscaling. Unless stated otherwise, we use 512 hidden features and 8 hidden layers. For the optimization we use the ADAM \cite{kingma2017adam} optimizer with learning rate $1 \times 10^{-4}$.

For the G-buffers, we provide all the information a traditional path tracer would require to evaluate the rendering equation of path tracing. The Rendering Equation gives outgoing radiance $L_o$: 

\begin{equation}\label{eq:rendering}
	L_o(\x, \w_o) = L_e(\x, \w_o) + \int\limits_\Omega L_i(\x, \w_i) \,\bsdf(\x, \w_o,\w_i)\,\cos\theta_i\, \mathrm{d}\w_i
\end{equation}
	
\noindent with $L_e$ the emitted radiance and $\theta_i$ the angle between the surface normal and the incoming direction $\omega_i$; We create first-intersection G-buffers with the world position of the intersection $\x$, normal of the surface $n$, reflectance and roughness of the BSDF $\bsdf$ and outgoing direction $\w_o$. The normal and material information help the network understand the existing correlations between these signals and outgoing radiance $L_o$.

We optimize the neural generator to map this input to the value of the integration over the hemisphere. Emission is also computed as a first-intersection buffer and is passed through to the output directly.

The world position $\x$ conditions all the other inputs since it is where the integration happens. For this reason we precondition the \NNtype to the position G-buffer by passing it alone through the first network layer. 
In subsequent layers all the buffers are concatenated with the global information of the scene representation vector and passed to the network; this is similar in spirit to NeRF~\cite{mildenhall2020nerf} that inputs only position to the first layers. 
We experimented with Fourier features~\cite{tancik2020fourfeat}, but this resulted in artifacts due to the noise in the training data.
We show the effect of this choice in Sec.~\ref{sec:evaluation}, Fig.~\ref{fig:ablations-precondition}.

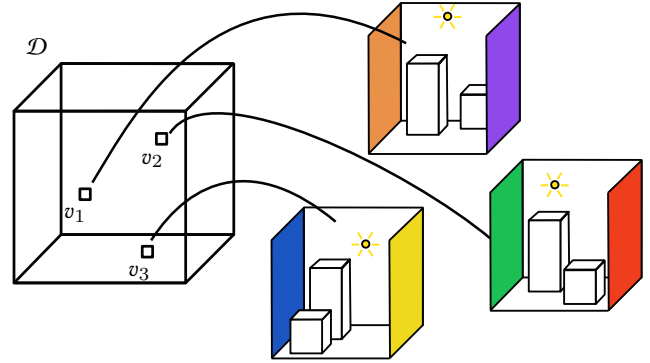
\begin{figure}[!t]
	\begin{overpic}[width=\linewidth]{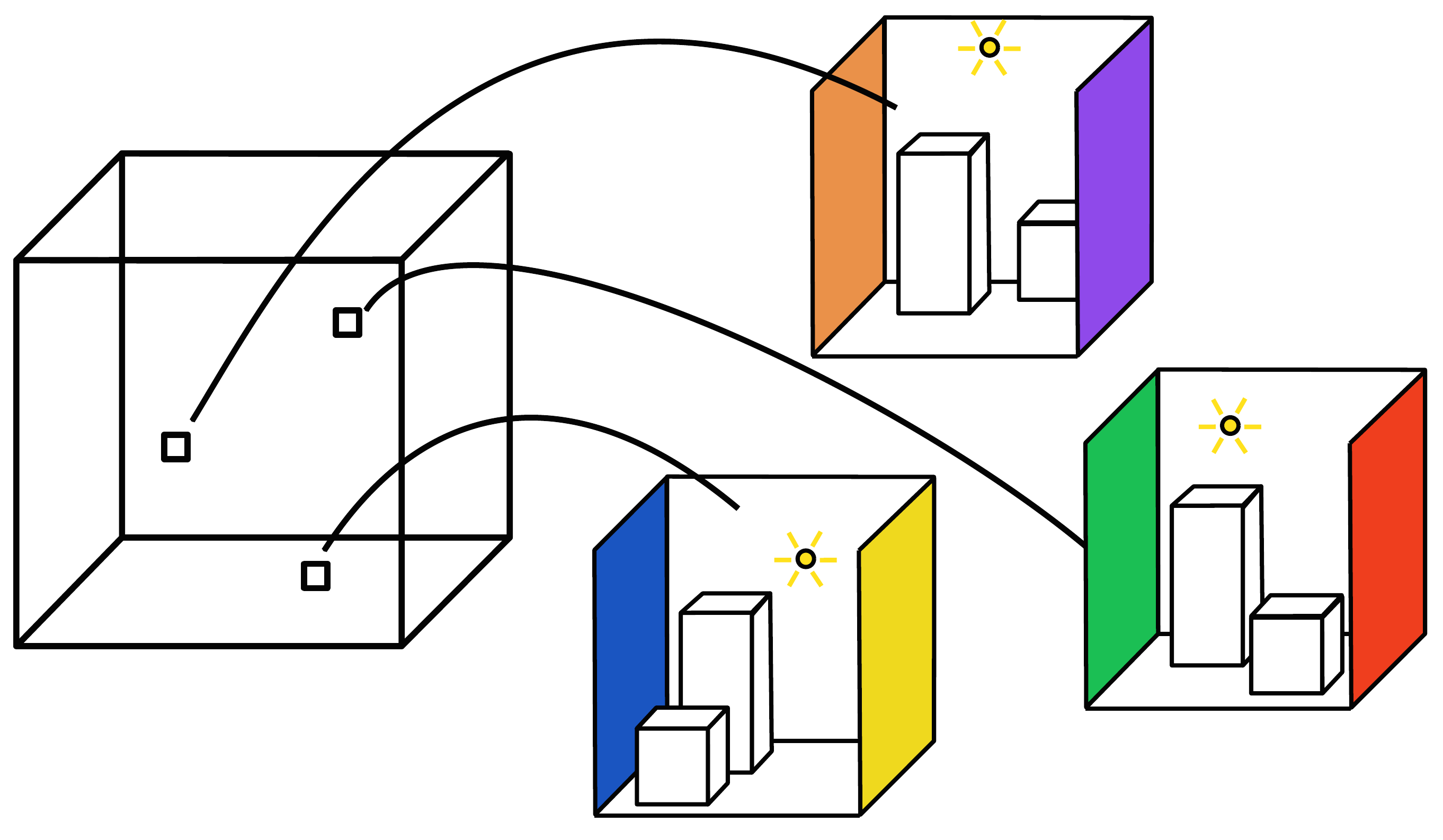}
		\put(3,49){$\mathcal{D}$}
		\put(9,23.5){$v_1$}
		\put(21,32){$v_2$}
		\put(19,14.5){$v_3$}
	\end{overpic}
	\caption{A point $v_i$ in the data space $\mathcal{D}$ defines a scene instance out of all the possible configurations of the \emph{variable} scene.}
	\label{fig:space}
\end{figure}

\section{Active Data Space Exploration}
\label{sec:active-exp}

For a given variable scene, we will optimize a neural generator using on-the-fly synthetic training data; we describe the training process and loss in Sec.~\ref{sec:training}. This training data is generated within the space $\mathcal{D}$ of all possible configurations of the scene. 

Each point in this space is defined by the values of the scene variables of the explicit scene representation vector $v$. Since the scene variables are normalized, the data space \D can be seen as a hypercube, see Fig.~\ref{fig:space}. 

A uniform random sampling of this high-dimensional space converges to a local minimum with low quality (see Sec.~\ref{sec:evaluation}).

\begin{figure}[!ht]
	\begin{overpic}[width=\linewidth]{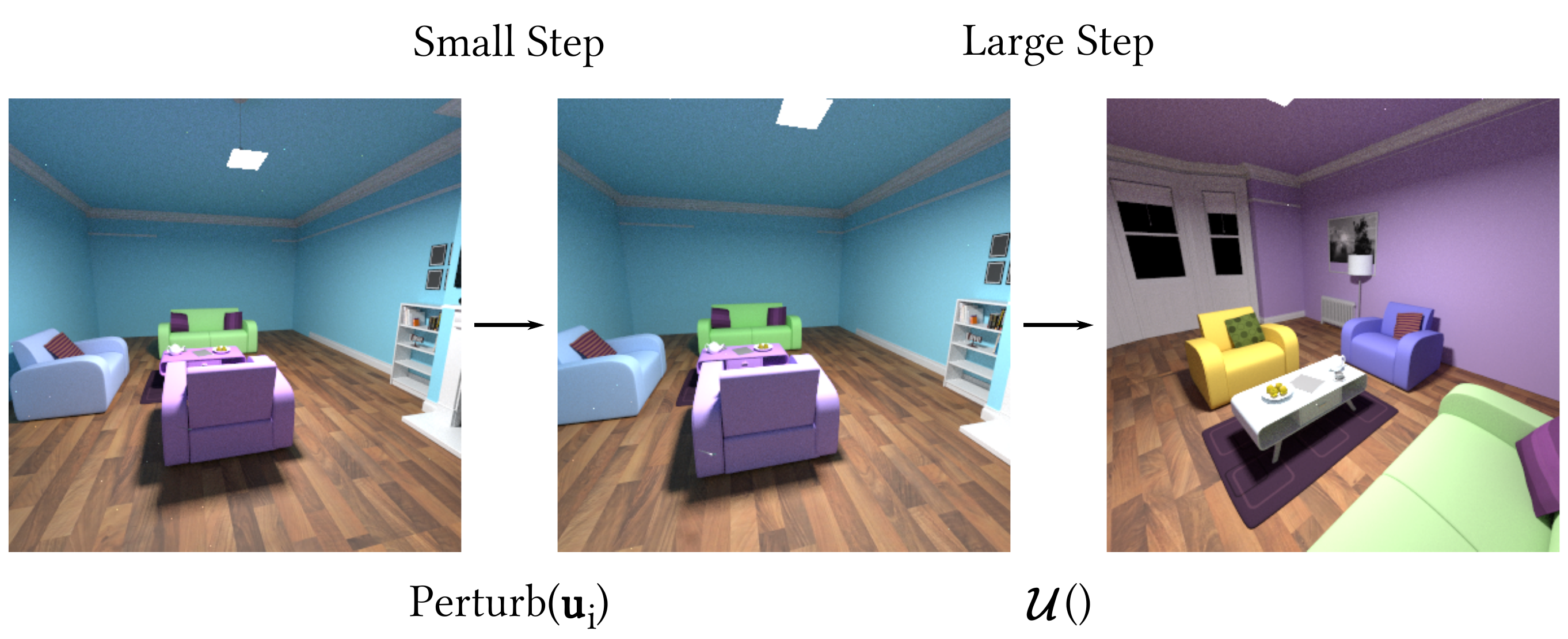}
	\end{overpic}
	\caption{ Visualization of the impact of a small and large step on a variable scene. In small steps (left) minor perturbations are applied -- here the light source, furniture positions and materials have been altered slightly. In large steps (right), major changes have been applied to the scene (position of furniture, albedo of objects etc.).}
\vspace{-2mm}
	\label{fig:largesmall}
\end{figure}

\begin{figure*}
	\centering
	\begin{overpic}[width=\linewidth]{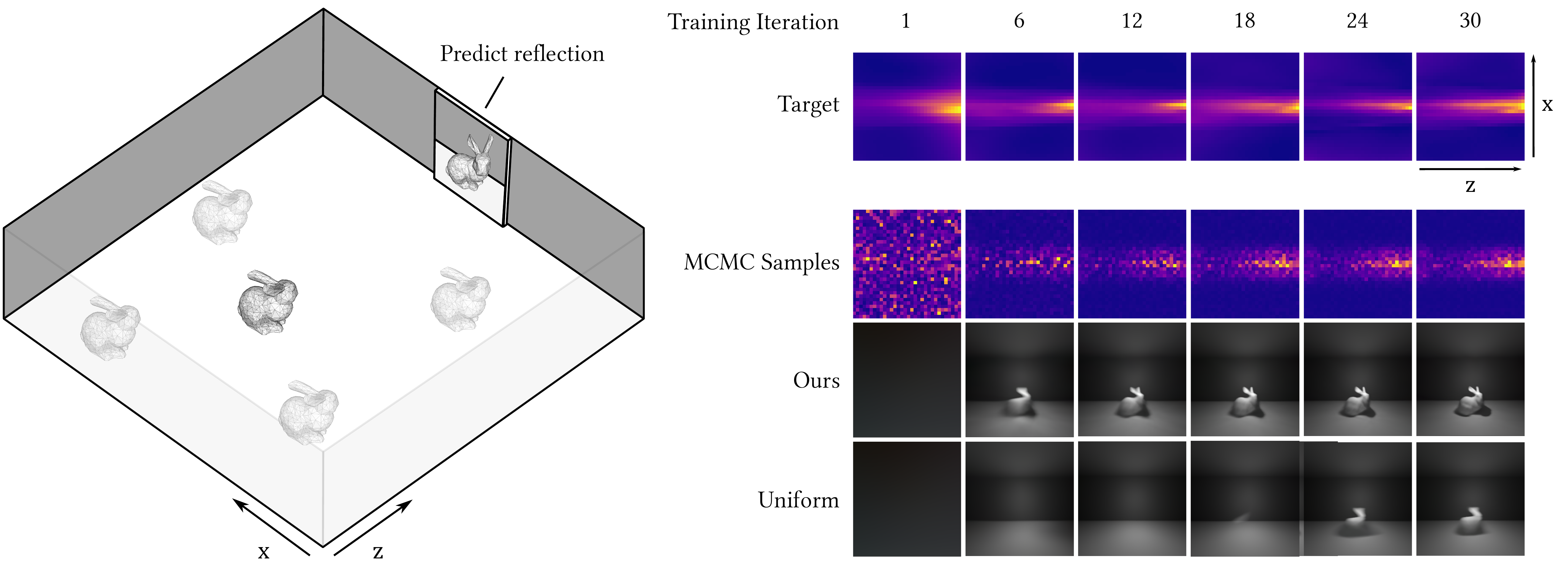}
	\end{overpic}
	\caption{ We test our method in this simple example to verify the convergence of the samples generated by Active Exploration. In this scene the variable parameters are the X-Z placement of a Bunny figure in an empty room (left). We fix the viewpoint to always look into the mirror and ask our generator to predict the reflection. The Bunny appears in the reflection for only a specific range of X values. We plot the target function (loss times gradients) for different X-Z values of the Bunny position (the heatmap can be seen as a top down view of the room) through training iterations, on the right. We show that the 2D histogram of Bunny placement from our Ative Exploration, after a burn in period, starts following the distribution of the target function. As a result the reflection of the Bunny starts appearing much sooner compared to uniform sample generation. }
	\label{fig:ablation-mcmc-distribution}
\end{figure*}

To overcome this difficulty, we propose \emph{Active Exploration} of the space $\mathcal{D}$ of training samples. The ability to generate on-the-fly training samples defined by the explicit vector $v$ offers great flexibility, allowing us to \emph{interleave} sample generation and training.

The high-level goal is to find a sampling strategy that will find samples in \D that maximize the progress of training and locally explore these pockets of importance.
We introduce a MCMC exploration strategy that guides sampling of $\mathcal{D}$, towards scene configurations where the network struggles to recreate global illumination. MCMC is well suited to searching such high-dimensional spaces, and has proven its utility both in learning~\cite{welling2011bayesian} and illumination~\cite{veach1997metropolis}.
	
The Markov Chain is initialized with a state picked uniformly from the hypercube of the data space $\textbf{u}=\textbf{u}_0\in\mathcal{D}$, i.e., a random configuration of the variable scene. The next proposed state is sampled from the proposal distribution $\textbf{v}\in T(\textbf{u}_i\rightarrow\textbf{v})$. 
Similar to the Primary Sample Space exploration \cite{kelemen2002simple} we balance global and local exploration of the space with large and small steps, by choosing large steps with probability $p_{\mathrm{LS}}$. Specifically:
	
	\begin{equation}\label{eq:proposal}
		T(\textbf{u}_i\rightarrow\textbf{v}) = \begin{cases}
			\mathcal{U}() \quad \textrm{ with probability }p_{\mathrm{LS}}=0.3 \\
			\textrm{Perturb}(\textbf{u}_i) \quad \textrm{else}	
					\end{cases}
	\end{equation}
\subsection{Markov Chain Exploration}
\label{sec:MCMC}

The data space \D can have arbitrarily high dimension, depending on how much variability exists in the scene. Our goal is to generate training samples that follow the distribution of sample importance, i.e., the impact of the sample on training. In MCMC terminology, our target function $f$ and corresponding target distribution $p$ should be defined so that the sampling process produces samples that maximize benefit for training.

The high dimensional space of \Dn, with pockets of importance, is ideal for a MCMC random walk exploration.
	
The hypercube of our data space $\mathcal{D}$ has a very similar structure to the primary sample space \cite{kelemen2002simple} and we take inspiration from the exploration choices of that method. 
The Metropolis-Hastings algorithm defines a proposal distribution $T(\textbf{u}_i\rightarrow\textbf{v})$ from a given state $\textbf{u}_i$ to a proposed state $\textbf{v}$. The target distribution $p$ is defined such that new states should be proposed and accepted for the Markov Chain to have a stationary distribution (i.e., the distribution at convergence) proportional to the target function. 

Our goal is to define a target distribution that will guide the training process to samples that accelerate training. Previous work has suggested different metrics of sample importance~\cite{zhao2015stochastic}. Two common such metrics are the training loss or the norm of the gradients after a backward pass. 
 
Our experiments showed that if only the loss is used, MCMC doesn't take into account where the network can improve the most. 

However, the \emph{product} of the loss and the norm works well, see Fig.~\ref{fig:ablations-target}.
Since we use ADAM~\cite{kingma2017adam}, instead of the norm of the gradients we use the norm of the total step to take into account the momentum and RMSprop \cite{Tieleman2012}.

The small step involves applying normally distributed perturbations to each component of $\textbf{u}_i$. A visualization of the impact of these steps on the final rendering is shown in Fig.~ \ref{fig:largesmall}. Since the proposal distribution is symmetric, meaning $T(\textbf{u}\rightarrow\textbf{v})=T(\textbf{v}\rightarrow\textbf{u})$, the acceptance probability of the proposed state similar to the Metropolis-Hastings algorithm is:
	
	\begin{equation}\label{eq:acceptance}
			\alpha(\textbf{u}_i\rightarrow\textbf{v})~=~min\left(1, \frac{p(\textbf{v})}{p(\textbf{u}_i)}\right)
	\end{equation}

\noindent The acceptance probability transforms the Markov Chain's stationary distribution to the target distribution. In our case we have a) an evolving target distribution that b) changes based on the samples we provide. 

In our experiments, the acceptance probability of Eq.~\ref{eq:acceptance} does not converge to the target distribution fast enough, i.e., before it has changed. For this special case we propose instead a more aggressive acceptance policy:
	
\begin{equation}\label{eq:outacceptance}
	\alpha(\textbf{u}_i\rightarrow\textbf{v})~=~\begin{cases}
	1 \quad	\textrm{if} ~~p(\textbf{v}) > p(\textbf{u}_i) \\
	0 \quad	\textrm{else}
	\end{cases}
\end{equation}
	
\noindent This acceptance probability has the desirable property that the more we remain in a state the more the target function -- which is related to the error -- decreases for this state. If we assume that the network can represent this state, then it will learn from it, meaning that the gradients and error will decrease, allowing a new proposed state to be accepted. \NEWTEXT{If there are states that cannot be represented (e.g., pixel perfect reflections) the gradients will guide the MCMC towards states that still have room for improvement avoiding the issue of getting stuck.}

In the initial phase of data generation, known in the literature as the burn-in phase, the Markov Chain the samples do not follow the target distribution. To alleviate this issue we use 16 Markov Chains in parallel, one for each patch rendered, leading to a shorter burn-in phase. This can be seen in Fig.~\ref{fig:ablation-mcmc-distribution} "MCMC Samples".

We evaluate our proposed acceptance policy and the sample distribution in a simple scenario shown in Figure \ref{fig:ablation-mcmc-distribution} and by disabling sample reuse. Here the 2D variable parameter is the position of a Bunny figure in a room with a mirror at the center of its wall. We task the generator with predicting the reflection (viewpoint is fixed to always look into the mirror). For this case the static reflection of the walls is learned quite easily but the variable reflection needs the Bunny to be placed in view of the mirror. Our method correctly does so and leads to its reflection appearing much faster compared to Uniform sampling.

\section{Training and Self-Tuning Sample Reuse}
\label{sec:training}

For training, we use the combination of $L_1$ and structural dissimilarity loss, as in Granskog et al. \shortcite{granskog2020compositional}. Since rendering is still slow it is beneficial to reuse samples as much as possible. We next discuss our self-tuning sample reuse and resolution enhancement methods.

\subsection{Self-tuning Sample Reuse}
\label{sec:reuse}
 
Traditional supervised deep learning typically uses a fix sized pre-computed dataset and runs optimization steps many times on batches, running through the entire dataset several times. Each such run is referred to as an epoch, resulting in the reuse of each data point many times.

In our case, we are generating training samples on-the-fly, and thus we do not have the notion of epochs. However, sample generation is costly (typically 2.5 sec for the 16 32x32 patches), and it is thus important to reuse training samples as much as possible, to speed up training, and also prevent the network from forgetting over the course of training. We do this by introducing a new self-tuning sample reuse strategy based on the divergence between the loss of newly seen data points and those already seen, to achieve a balance between overfitting and training speed.

Inspired by these observations, we achieve this balance by tracking two different losses $\LN$ and $\LE$, i.e., the loss of newly generated, unseen samples and the loss of the previously generated samples that were already used to train the network. Both are tracked using an exponential moving average to lessen the effect of the stochasticity of the optimization process. 
When $\LE$ starts decreasing faster than $\LN$, thus diverging from it, our model is starting to over-fit (as new data is performing worse than previously generated data). In this case we need new samples to augment the size of our dataset. This can be done by decreasing $p_s$, i.e., reusing with a lower probability.

We start training on-the-fly, generating and storing a new sample for the first 100 samples. After this short initialization, for each new step we randomly decide to reuse a previously generated sample, or generate and store a new one. The decision is made based on a Bernoulli distribution with a self-tuning probability $p_s$ over steps $s$, representing the probability of reusing a training sample.

We build probability $p_s$ to satisfy two goals. First we would like $p_s$ to be as high as possible, so that we save as much computation as possible. But if it is too high, or even equal to 1, we would over-fit to the already generated samples and stop exploring the space of parameters. Thus $p_s$ should also be sufficiently low to avoid over-fitting. Over-fitting is usually measured by the difference of performance of a model between a training and validation dataset.

We propose a mechanism with a single parameter to control $p_s$:
\begin{equation}
p_s=\sigma(\LE-\LN+\beta)
\end{equation}

\noindent where $\sigma$ represents the sigmoid function and $\beta$ is the parameter controlling the reuse probability when both losses are equal. This formulation decreases the probability of reusing a sample when $\LE$ is lower than $\LN$. \NEWTEXT{Intuitively the above equation is derived by associating the losses to negative log-likelihood of probability distributions parameterized by the ground truth. More details can be found in the supplemental materials.} Since one component of the MCMC target function maximizes the loss (see Sec.~\ref{sec:MCMC}) we use only large step samples to keep track of both $\LE$ and $\LN$. In all our experiments 
$\beta$ is set to 4.6, to have $p_s=0.99$ when $\LN=\LE$. When a sample is reused we build a batch of training images from the stored patches. We use the previous loss of the sample as a weight, i.e., setting the probability of selecting a patch proportional to its last recorded loss. We update the weight of a sample whenever it is reused, using the network loss on that sample in the current iteration. This allows hard samples to be reused more often and discards those for which the network performs well, leading to better adequacy between reuse and MCMC.

\subsection{Resolution of Training Images}
\label{sec:resolution}

One of the main advantages of using a \NNtype for the generator architecture, as demonstrated by \cite{granskog2020compositional}, is its performance during inference on much higher resolutions than that used for training. Shading effects that depend heavily on G-buffers such as textured diffuse materials benefit from buffer upscaling, providing improved quality. That is less true for high frequency view dependent effects, such as reflections, that are are typically small, and band-limited by the resolution of training images. Our goal is to progressively reduce the area each training sample covers, allowing the model to gradually focus on such effects.

\begin{figure}[!h]
	\includegraphics[width=\linewidth]{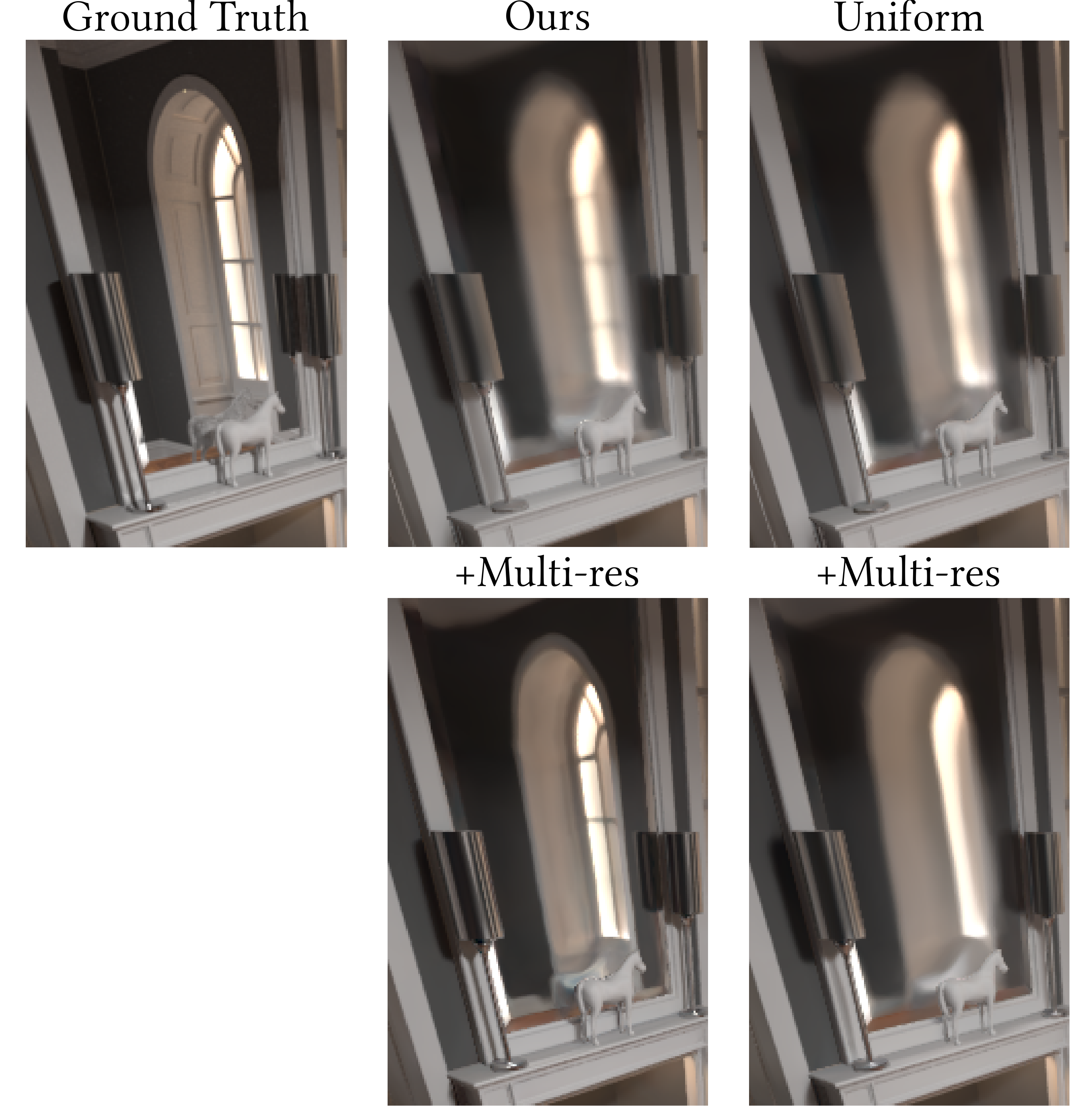}
	\caption{\label{fig:ablations-resolution}
		\NEW{New GTs, reorganized.} Ablation study for adaptive resolution MCMC vs Uniform training. First row: the resolution is always 128x128. Second row: we progressively focus resolution, improving quality.}
\end{figure}

We adopt a multi-resolution approach to address this. We start training with 32x32 patches extracted from 128 by 128 pixel images with a $90^\circ$ field of view. Note that the MCMC controls the respective patch position on the image plane and that we only render the patch pixels. We then progressively increase the resolution of the images used to select the 32x32 patches; we found that doing so by 4 pixels every 2000 iterations worked well, all the way up to 600 by 600 which is closer to our target resolution. This shrinks the area of the patch on the sensor and allows the network to observe finer details in hard regions, such as reflections, during training. This process is made possible by the MCMC exploration, due to its ability to locally explore the scene configuration through the small steps, and to adapt to this progressive change in resolution.
On the other hand, adopting such a multi-resolution approach with uniform sampling of $\mathcal{D}$ results in worse overall results as it decreases the probability of observing a given point in the scene. \NEWTEXT{This makes} sampling even less efficient (see Fig.~\ref{fig:ablations-resolution}) resulting in lower perceived image quality.

\begin{figure*}[!hp]
\begin{overpic}[width=0.9\linewidth]{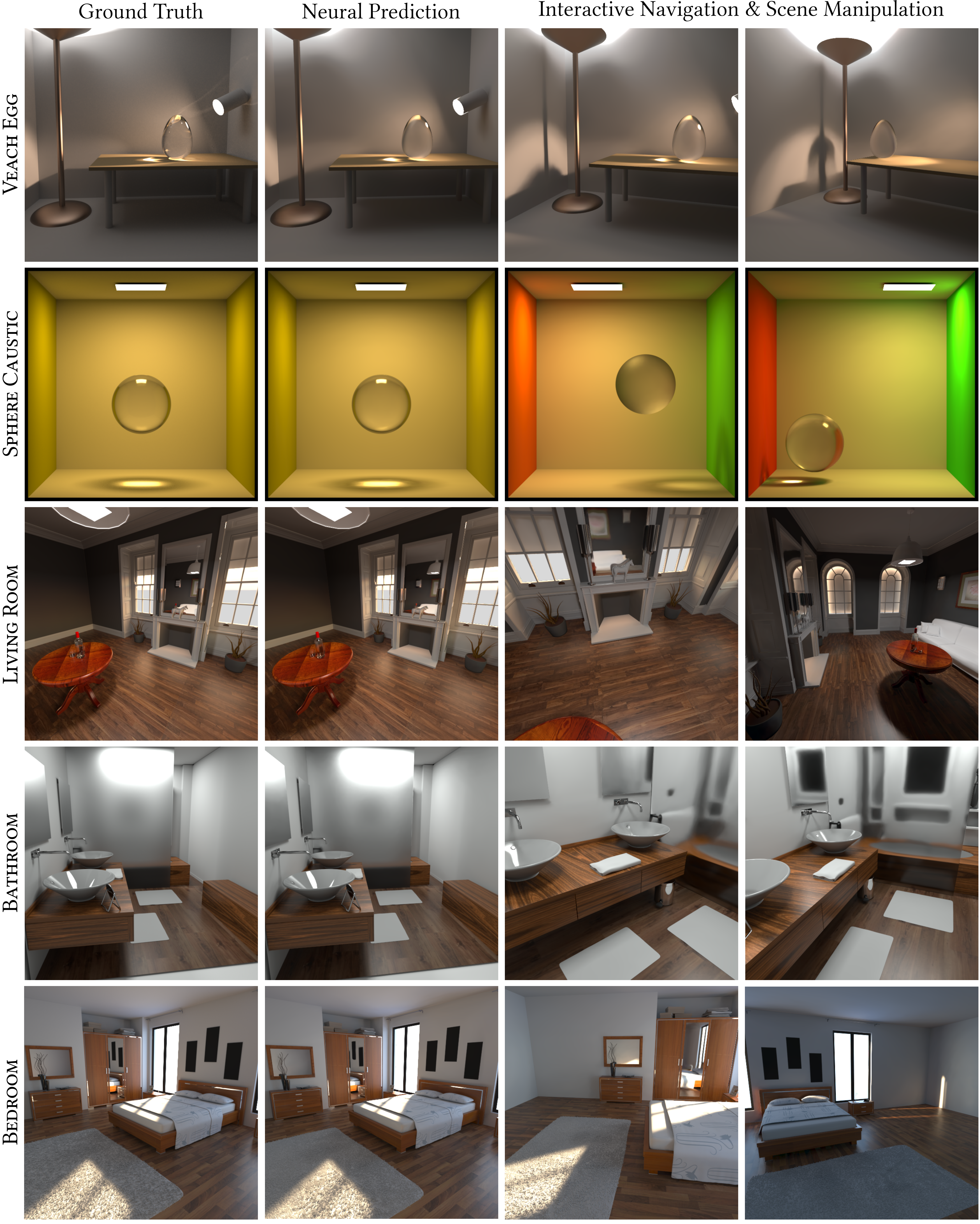}
\end{overpic}
\caption{\label{fig:resultspage}
	\NEW{Updated GTs, added metrics.} Results of our method for 5 different scenes each with different variations. Note how we can capture view changes (all but row 2) reflections, approximate caustics, global illumination etc. all at interactive rates. The scene variables include: material albedos and roughness (\textsc{Sphere Caustic}, \textsc{Bathroom}), moving and rotating objects (\textsc{Veach Door}, \textsc{Living Room}) and time of day \textsc{Bedroom}. Please also see supplemental videos.
\vspace{-1.5mm}
}
\end{figure*}

We show results for several variable scenes; please see the supplemental videos to best appreciate the quality of our results in these dynamic scenes. We also present comparisons to previous work and analyze the various design decisions of our solution through ablations studies and quantitative evaluation.

We have implemented our system in Python interfaced to Mitsuba 2 ~\cite{nimier2019mitsuba} which we use to render global illumination and G-buffers. We use between 200 and 24,000 samples per pixel for ground truth renderings, depending on the scene see Table~\ref{tab:gt-spp}.

We allow transformation of geometry and lights, material editing and viewpoint changes, including discrete events (e.g., changing between different materials, objects appearing/disappearing). 

Our prototype implementation runs at \OURFPS~ at inference/rendering time (900x900 resolution on a NVIDIA 3090 GPU), allowing interactive exploration of dynamic global illumination in variable scenes with potential applications in architecture, design, games, etc. 

Currently we only show results with a forward path tracer (the only integrator available in Mitsuba 2). However, our method is agnostic to the type of integrator and if we used a different renderer, we could train with bi-directional path tracing, Metropolis or any other method.

\begin{figure*}[!h]
\begin{overpic}[width=\linewidth]{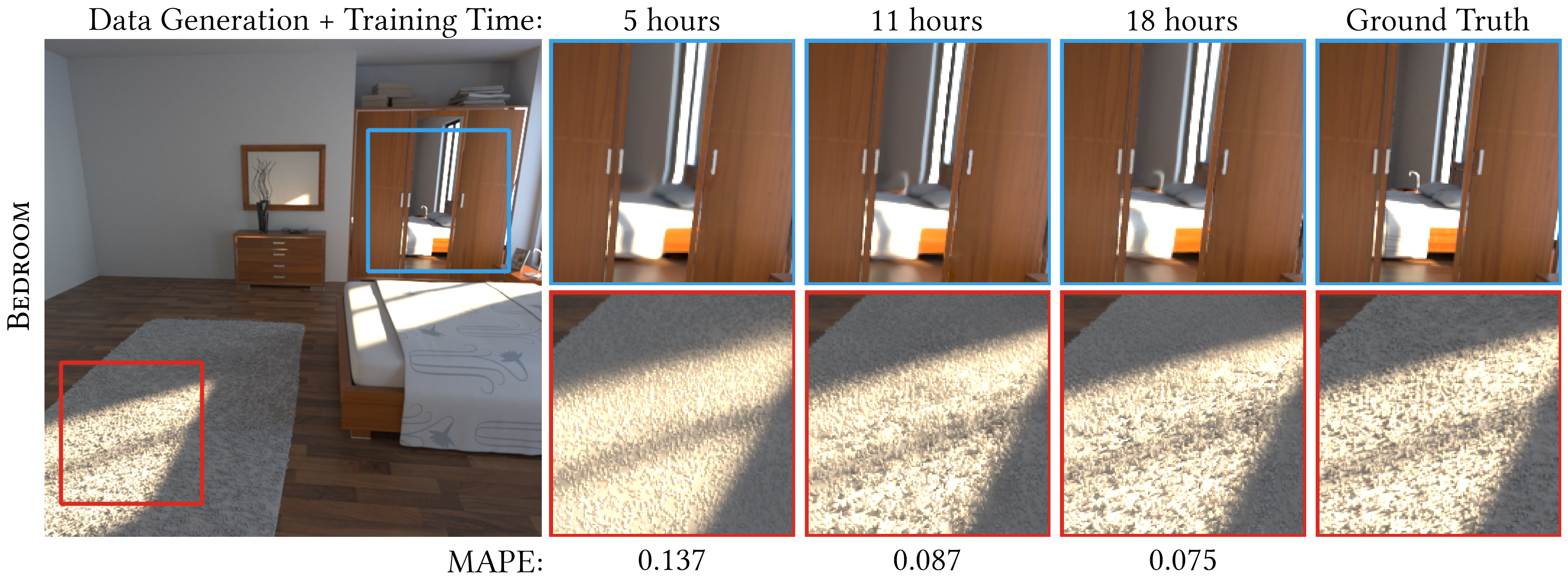}
\end{overpic}
\caption{
\label{fig:ablation-training-time}
\NEW{New Ground Truth.} Results of our method after increasing hours of training. Depending on the application if speed is valued over quality, our method yields plausible results after a few hours of training and data generation. For the best quality possible our method requires around 18 hours in the \textsc{Bedroom} scene.
}
\end{figure*}

\begin{figure}[!b]
	\begin{overpic}[width=\linewidth]{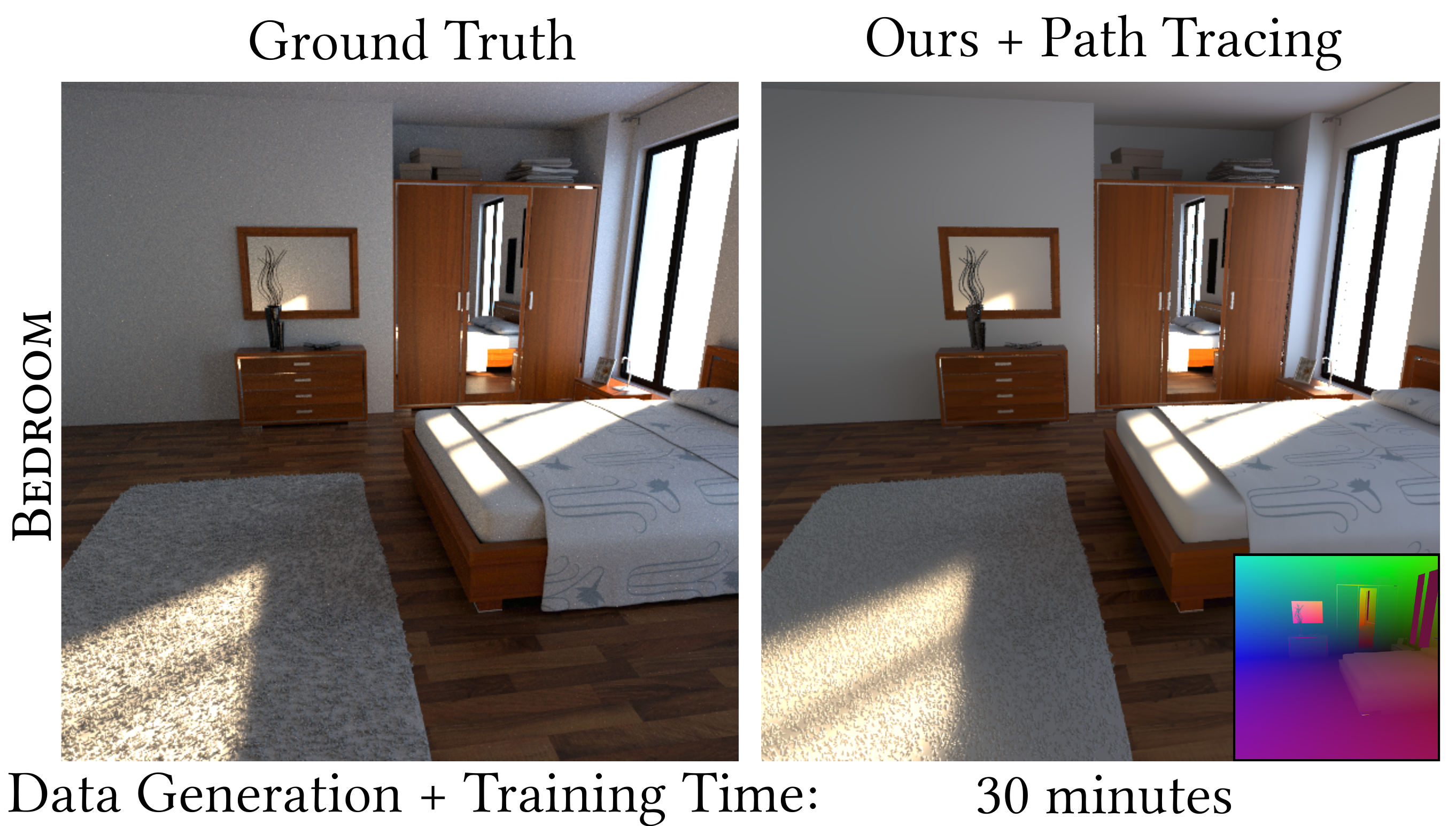}
	\end{overpic}
	\caption{
		\label{fig:ablation-specular}
		When ray tracing hardware is available our method benefits by tracing all specular bounces during the buffer generation (positions buffer in inset). This means that with only 30 minutes of data generation and training our method learns the non specular shading for the \textsc{Bedroom} scene. The harder high frequency details on the carpet still need full training to appear.
	}
\end{figure}

\begin{figure*}[!h]
\includegraphics[width=\linewidth]{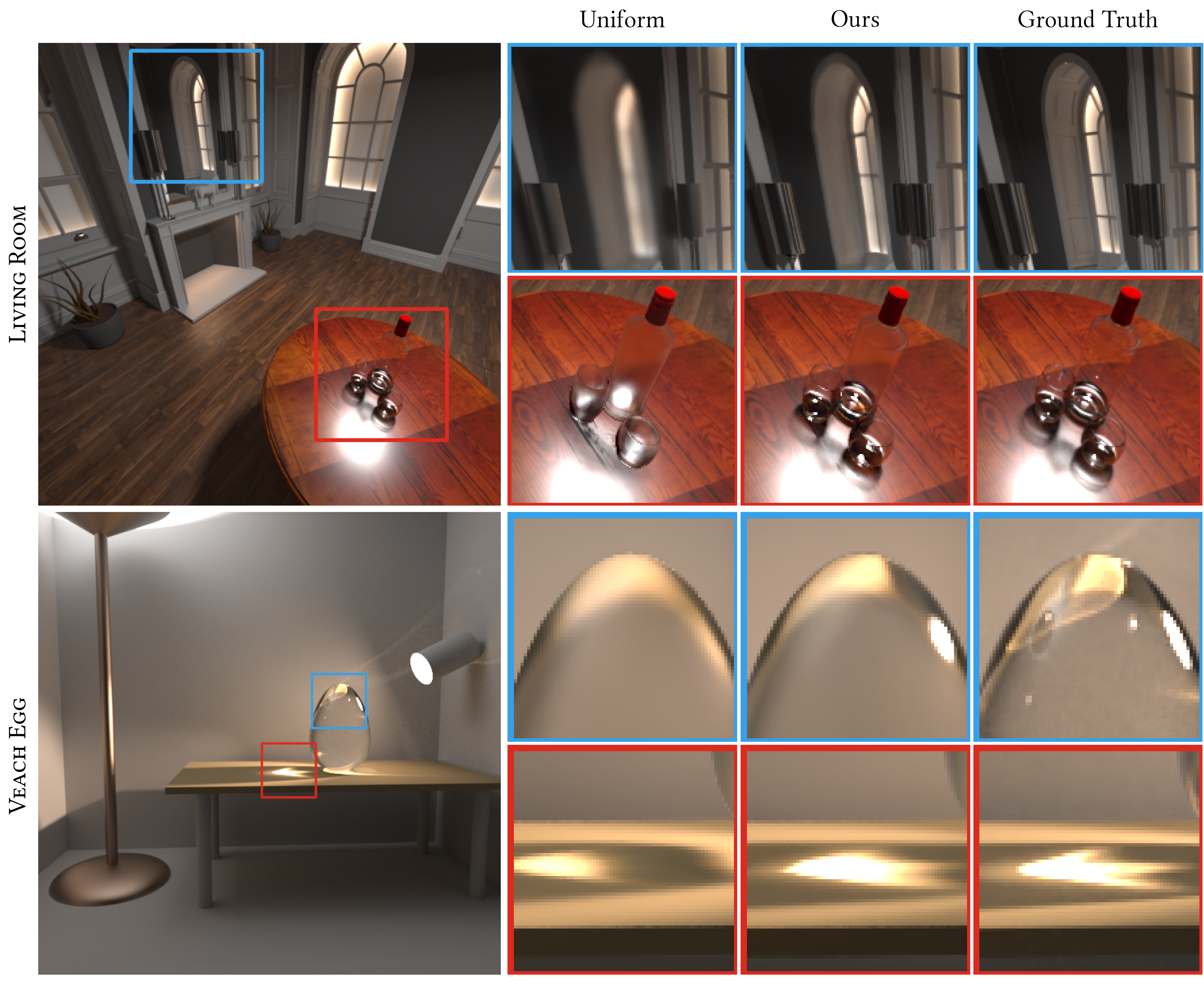}
\caption{\label{fig:ablations-mcmc}
\NEW{Converged GTs.} We compare our active exploration MCMC method vs. Uniform sampling of the space \D trained for the same time. We see that Uniform search running for the same time cannot produce sharp shadows, reflections and caustics.}
\end{figure*}

\section{Results, Analysis and Comparisons}
\label{sec:results}

\subsection{Results}

We present results of our method on several modified scenes from the Bitterli dataset~\cite{resources16}; the viewpoint is variable in all 7 scenes except \textsc{Sphere Caustic}, Figures \ref{fig:teaser} and \ref{fig:resultspage}. 

\begin{figure*}[!h]
	\centering
	\begin{overpic}[width=\linewidth]{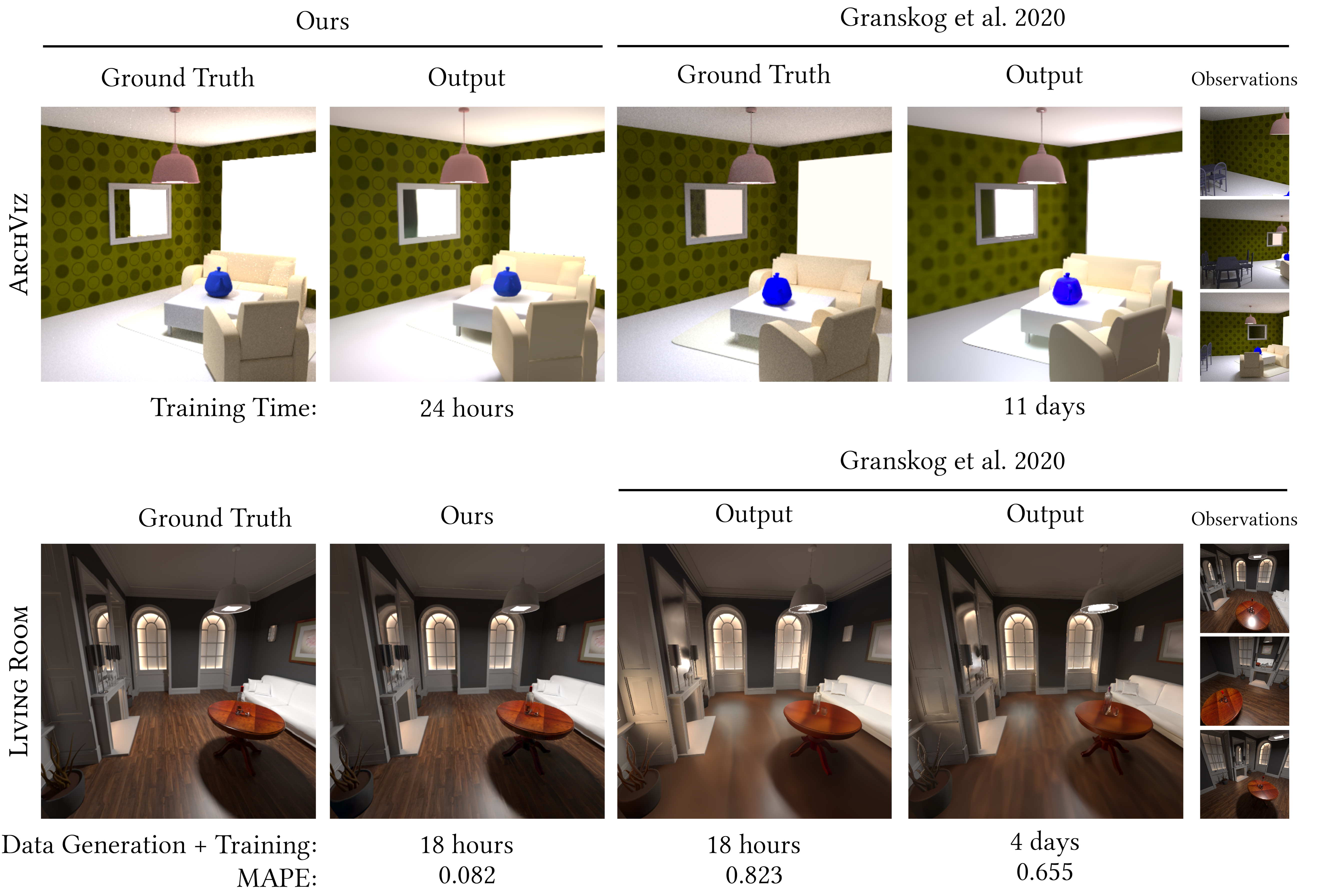}
	\end{overpic}
	\caption{\NEW{Added a 4 days version of Granskog et al. 2020 and MAPE metric for same time}Same quality (top) and same time (bottom) comparison with Granskog et al.~\shortcite{granskog2020compositional}. We show \NEWTEXT{result of} \textsc{ArchViz} for same quality \NEWTEXT{as ours} using the pretrained network provided by the authors in their rendering framework. \NEWTEXT{We also show} \textsc{Living Room} for same time \NEWTEXT{as ours} by training their method on our variable scene in our rendering framework. The 3 path traced observations required by Granskog et al.~\shortcite{granskog2020compositional} are shown on the right in both cases.}
	\label{fig:comp-granskog}
\end{figure*}

For the \textsc{Bathroom} scene, we added a showerdoor with variable roughness; additional variables are the intensity and position of the light source (total 8 dimensions). For the \textsc{Living Room} scene, we added blinds on the windows that can open and close; additional variables include the light intensity (7 dimensions). For the \textsc{Bedroom} scene, we have simulated variable sunlight with a distant source coming in through the window (6 dimensions). We present a modified version of the \textsc{Veach Door} scene, where the 
variable is the opening door (6 dimensions). We also have a modified Cornell Box, \textsc{Sphere Caustic} with variable wall colors and a large sphere that can move in the scene and vary in roughness, for a fixed viewpoint (11 dimensions). The \textsc{Spaceship} scene contains 3 variable emitters, 2 on the ceiling and one in the cockpit and variable viewpoint (8 dimensions). Finally in the \textsc{Veach Egg} scene we can vary the position of the glass egg and the spotlight emitter (9 dimensions). We show several configurations of each scene in Fig.~\ref{fig:resultspage} and Fig.~\ref{fig:teaser} and a path with variations in supplemental.
We train the scenes for 5-18 hours on a single NVIDIA RTX 6000. If training speed is important, we obtain a reasonable first approximation after a few hours, but longer training is required if we want to be very close to ground truth (see Fig.~\ref{fig:ablation-training-time}). 

Different application scenarios have different hardware resources. If the target platform is a modern desktop computer with a ray tracing GPU, specular interactions can be traced and not inferred. In this case our method needs less than an hour for acceptable results \NEWTEXT{more training is required to achieve the} highest possible quality as seen in Fig.~\ref{fig:ablation-specular}.

\begin{table}[!h]
	\includegraphics[width=\linewidth]{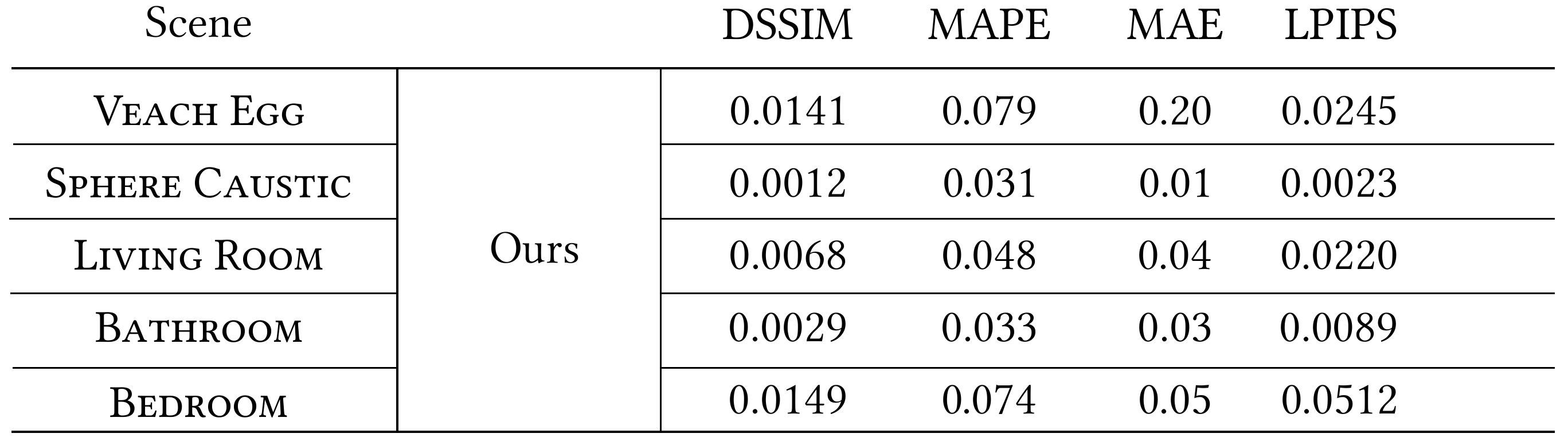}
	\caption{\label{tab:results-table} \NEWTEXT{Quantitative results using 4 metrics for the configuration shown in Figure~\ref{fig:resultspage}.}}
\end{table}

Our solution shows good temporal stability (see supplemental and video). The results show that we can capture a wide variety of light transport effects: global illumination (\textsc{Living Room} with different blind positions; Fig.~\ref{fig:resultspage}), soft shadows, glossy (or even partially specular) reflections (\textsc{Living Room}, transmission (\textsc{Bathroom}), caustics, (\textsc{Sphere Caustic, Spaceship, Veach Egg}) etc. A major strength of our approach is that we can render very hard light paths with good quality at interactive rates, e.g., the caustic in \textsc{Spaceship} (Fig.~\ref{fig:teaser}, or the shadow from the caustic in \textsc{Veach Egg}, Fig.~\ref{fig:comp-denoiser}, last row).
\NEWTEXT{The quantitative results in Tab. \ref{tab:results-table} show that we achieve low error rates in all scenes.}

\subsection{Comparisons}

The most significant comparison we will present is to Uniform sampling, since this clearly reveals the advantage of our active exploration approach. We also compare to Compositional Neural Scene Representations (CNSR) \cite{granskog2020compositional}, since we share similar inputs and some goals. The comparison mainly shows the benefits of our Active Exploration, explicit scene representation, and sample reuse in terms of training and inference speed. Finally a compelling alternative to our method for real time rendering of dynamic scenes is Real Time Path Tracing plus denoising. We compare with the state of the art denoising method of~\cite{isik2021anf}, further illustrating that our method is one of most efficient solutions for interactive rendering of hard light transport configurations, that require a very high sampling rate to be captured by path-tracing.

\begin{table}
\caption{\label{tab:gt-spp}
		Samples per pixel used for each scene during training.}
\setlength{\tabcolsep}{1pt}
\begin{tabular}{l| ccccccc}
 &
\small{\textsc{Sphere }} & 
\small{\textsc{Living }}  &
\small{\textsc{Bed}}  &
\small{\textsc{Veach }}  &
\small{\textsc{Bath }}  &
\small{\textsc{Space}}  &
\small{\textsc{Veach }} \\
 Scene &
\small{\textsc{Caustic}} & 
\small{\textsc{Room}}  &
\small{\textsc{room}}  &
\small{\textsc{Door}}  &
\small{\textsc{room }} &
\small{\textsc{ship}}  &
\small{\textsc{Egg}} \\
\toprule
spp & 200 & 400   & 400   & 600   & 800  & 1200   & 24000   \\
\end{tabular}
\end{table}

\begin{table}[!h]
	\includegraphics[width=\linewidth]{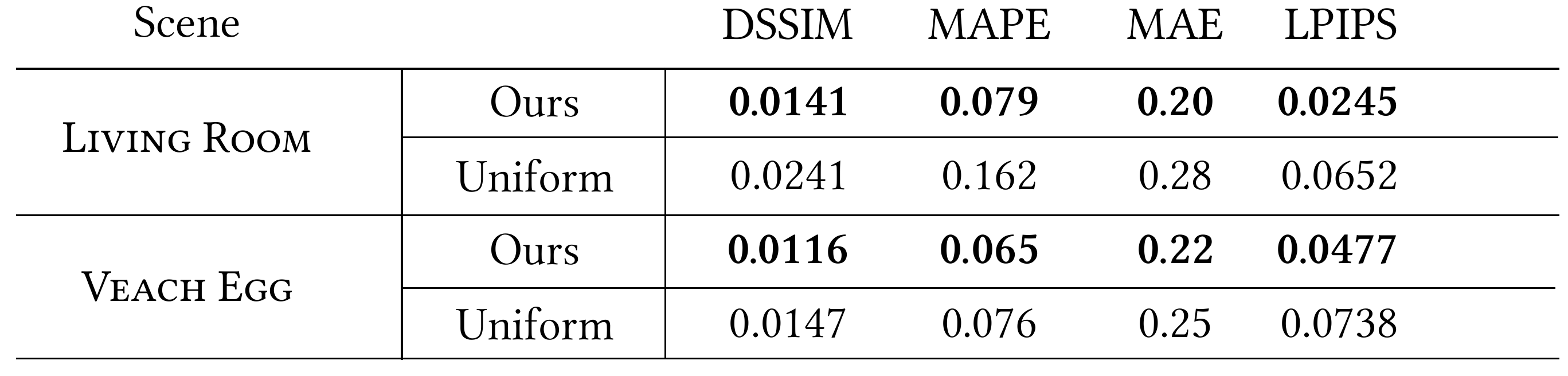}
	\caption{\label{tab:mcmc-table} \NEWTEXT{Quantitative results using 4 metrics for the configuration shown in Figure~\ref{fig:ablations-mcmc}.}}
\end{table}

\subsubsection{Comparison to Uniform sampling.}
\label{sec:comp-uni}

To evaluate the effect of MCMC active exploration
our first comparison is to a simple uniform sampling baseline (in Fig.~\ref{fig:ablations-mcmc}). To simulate uniform sampling, we replace our MCMC method with large steps only, that are always accepted; note that this baseline includes our sample reuse method, but not resolution adaptation since it gives worse results (Sec.~\ref{sec:evaluation}). As we can see, for the same computation time, active exploration achieves sharper reflections, caustics and shadows, thanks to the guiding sampling it affords. We tried to obtain equal quality with the uniform sampling, however this naive approach converges to a low quality local minimum. The results shown in Fig.~\ref{fig:ablations-mcmc} were generated with the best quality this approach could achieve; after this point in training the loss does not decrease. The videos in supplemental showcase the improvement Active Exploration provides in all different configurations of the scenes, e.g., the caustics in \textsc{Spaceship} and \textsc{Veach Egg}, glossy effects in \textsc{Bathroom}, and reflections in \textsc{Bedroom} or \textsc{Living Room}. In all cases, our method provides sharper results, generally much closer to the ground truth.
\NEWTEXT{This is confirmed with quantitative analysis in Tab.~\ref{tab:mcmc-table}.}

\begin{figure}[!h]
	\includegraphics[width=\linewidth]{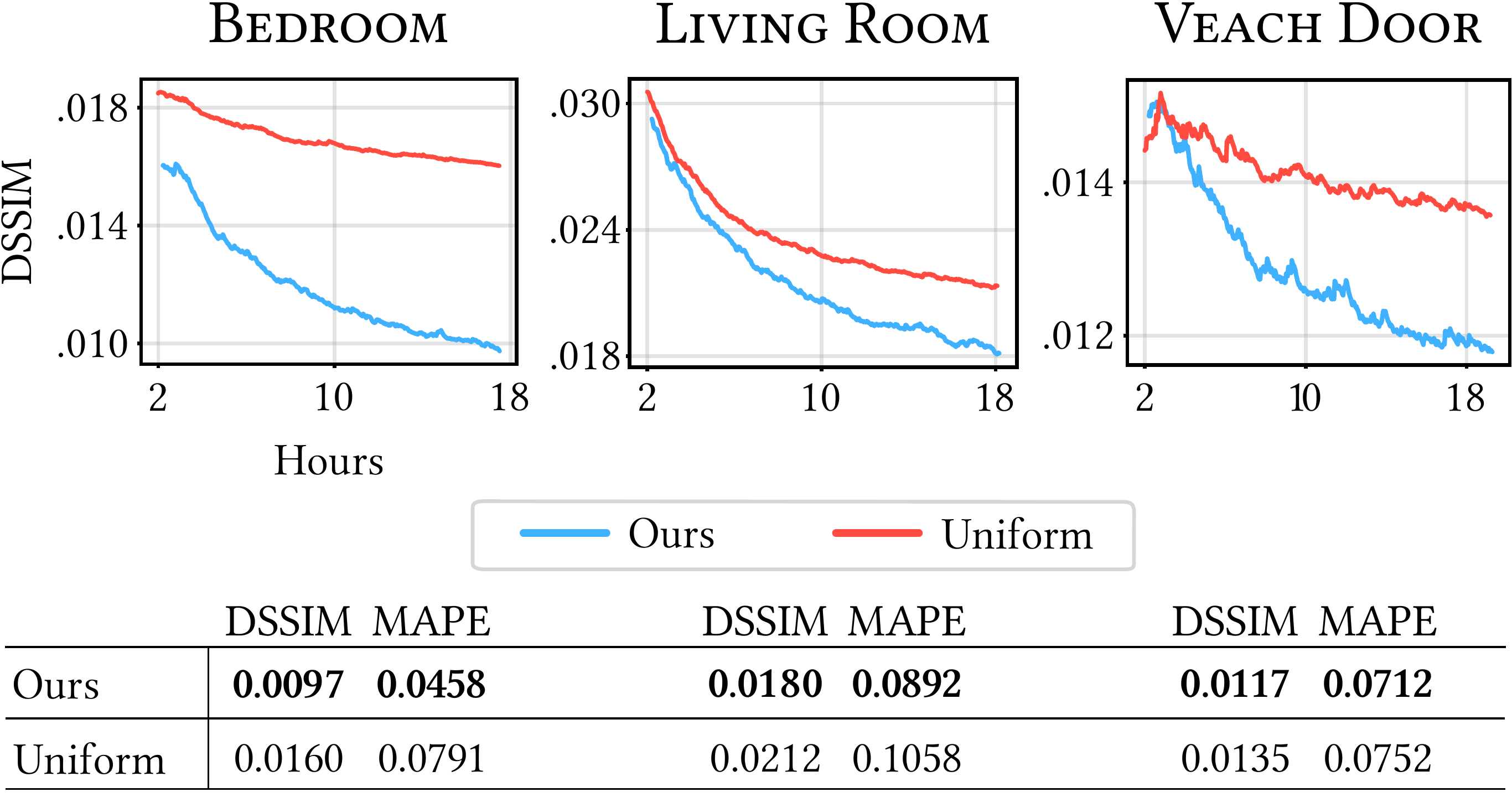}
	\caption{\label{fig:error} Quantitative evaluation of our method and ablations compared to ground truth (graphs start at 2h of training.)  }
\end{figure}

We also show quantitative results in Fig.~\ref{fig:error} using the Mean Absolute Percentage Error (MAPE), \NEWTEXT{DSSIM~\cite{loza2006structural}, Mean Absolute Error (MAE) and LPIPS~\cite{zhang2018unreasonable} error metrics} and a graph with the evolution of error over time. Since our main goal is to handle difficult lighting configurations, we select 10 frames from each path of each scene which correspond to such cases, and evaluate our method against ground truth; we show the selected frames for each scene in supplemental. 

\begin{figure*}[!hbt]
\includegraphics[width=\linewidth]{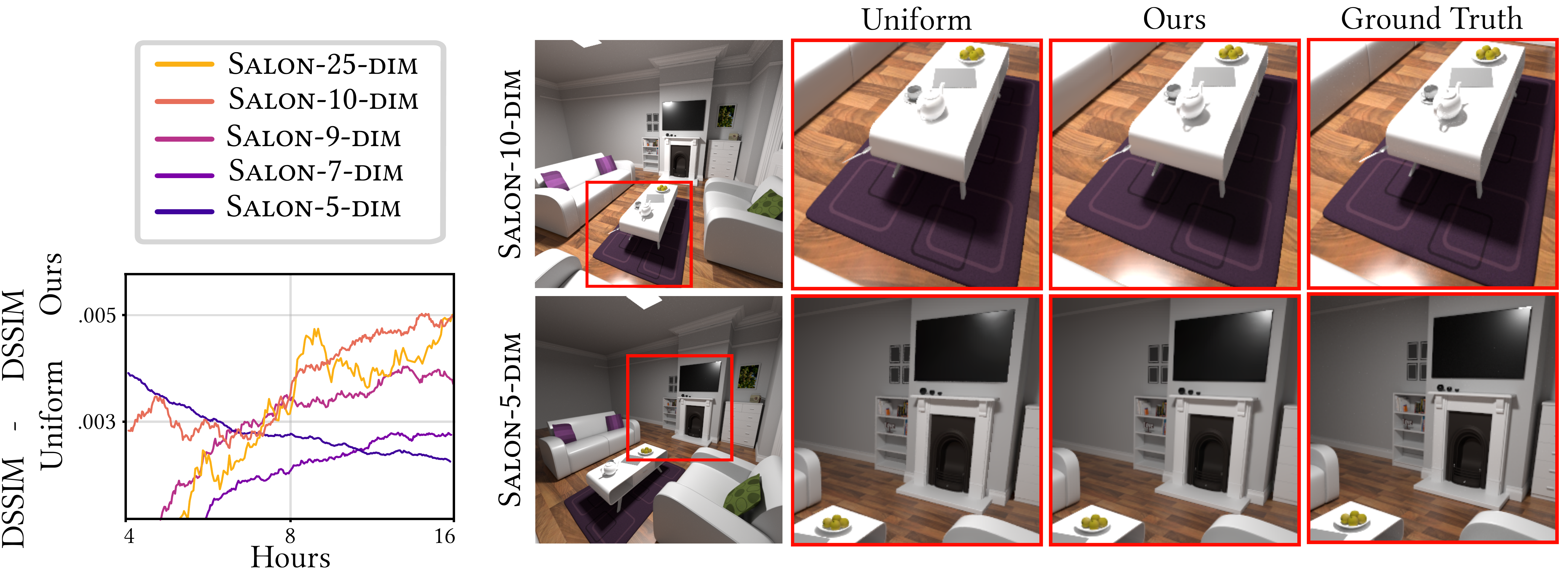}
\caption{\label{fig:variable-living}
\NEW{Updated figure and added insets.}Study of the number of dimensions on the effectiveness of our approach on the \textsc{Salon} scene. Left: we show the difference between the loss using the uniform approach and our method; the graph starts at 4h of training. In general, the benefit of our method increases with the number and complexity of the variable elements in the scene, but some elements are more important: despite going from 10 to 25 dimensions the difference of gain between \textsc{Salon-10-dim} and \textsc{Salon-25-dim} -- which only involves albedo changes -- is smaller than adding a single important dimension such as light position (difference from \textsc{Salon-7-dim} to \textsc{Salon-9-dim}. Right, for low dimensions (bottom row, \textsc{Salon-5-dim}) our method slightly improves the glossy highlight on the tv compared to uniform sampling; however, once the dimensions increase (top row, \textsc{Salon-10-dim}), we capture a lot of effects completely missed by the uniform, namely the tv and floor glossy highlight as well as the detailed shadows of the teapot on the table.
}
\end{figure*}

\subsubsection{Comparison to CNSR}

We compare our method to Compositional Neural Scene Representations~\cite{granskog2020compositional} (CNSR) using the variable \textsc{ArchViz} scene, the more complex of the two datasets used in CNSR; Our implementation of this scene has 71 dimensions. For best-effort same quality comparison we use a pretrained model provided by the authors. Note that the ArchViz dataset "consists of variations of a living room with a dining area" \cite{granskog2020compositional}. Both the pretrained model of Granskog et al.~\shortcite{granskog2020compositional} and ours are trained on identical data involving variations of this scene. We recreated the ArchViz variable scene in our framework as closely as possible, using publicly available resources~\cite{granskog2020compositional}. The CNSR pretrained model is trained on a dataset of 9000 sample points. Each point includes 16 batches of 3 observations at 64x64 resolution and a query image at the same resolution, trained for 1M iterations. 

The high complexity of this scenes' variations (constrained, specific configurations, e.g., the teapot appears at a specific position on the table etc.) challenges our base method; we show results using 256 features/layer and without resolution enhancement. This gives blurrier results (on a par with Granskog et al. in terms of quality) but avoids high frequency artifacts. Our method achieves same qualitative results with \emph{36 hours} of \emph{both training and rendering}. In comparison Granskog et al.~\shortcite{granskog2020compositional} needs \emph{11 days of only training} (accounting for hardware differences), and an unspecified amount of rendering time to generate the data.

We also retrain CNSR on three of our scenes, providing same time comparisons on \textsc{Living Room}, \textsc{Bedroom} and \textsc{Veach Door}. For this we used the publicly available code provided by the authors to train on data generated by our framework. As in the case of the \textsc{ArchViz} scene we use 16 batches of 3 observations at 64x64 resolution and a query image at the same resolution to train their model. 

The results of the same quality \textsc{ArchViz} and same time \textsc{Living Room} comparisons with Granskog et al.~\shortcite{granskog2020compositional} are shown in Fig.~\ref{fig:comp-granskog}. Additional examples are shown in the supplementary material. Our method achieves much sharper results that are significantly closer to the ground truth.

We want to note that this comparison is provided only as an indication of the efficiency of our approach, since the goals of the two methods differ in several ways.

\subsubsection{Comparison to ANF}
\label{sec:comp-anf}

\begin{figure*}[!h]
	\centering
	\begin{overpic}[width=\linewidth]{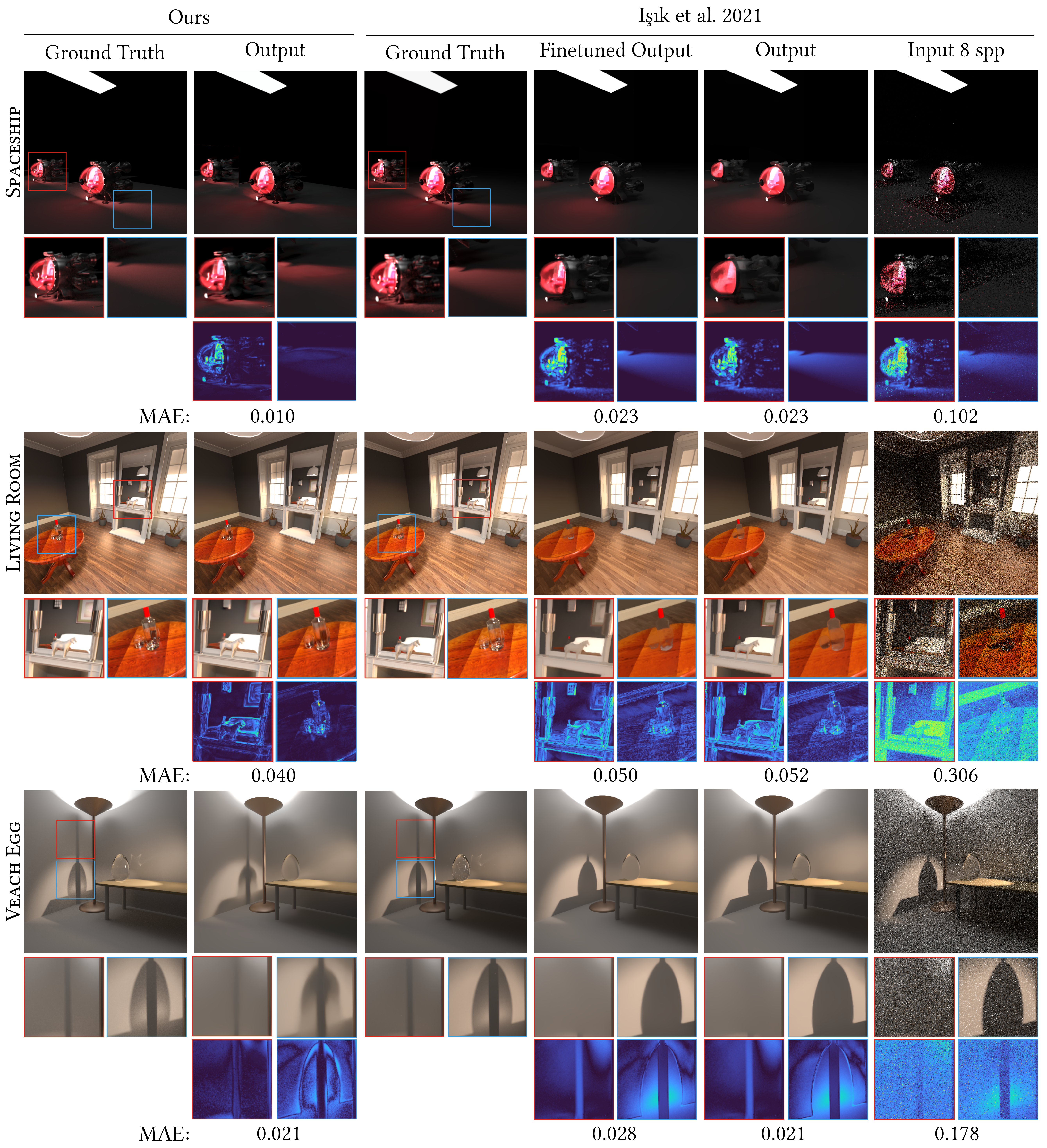}
	\end{overpic}
	\caption{ Same time comparison with Işık et al.~\shortcite{isik2021anf}, fine-tuned on our scenes. Note how our neural renderer captures hard light paths, e.g., caustics (\textsc{Spaceship}) or even shadows from caustics (\textsc{Veach Egg}) that are almost completely missing from the path-tracing + denoising solution.}
	\label{fig:comp-denoiser}
\end{figure*}

We compare with the recent Affinity of Neural Features (ANF) denoising method~\cite{isik2021anf} in Figure \ref{fig:comp-denoiser}. For a fair comparison we take the pretrained model provided by the authors and fine tune it in each specific scene, using the authors original implementation. 
Since our method uses a different renderer than ANF (Mitsuba 2 vs PBRT v3), we give the same budget in terms of pixels generated during fine tuning.
Also we fine tune the pretrained ANF model using sequences of 8 frames in random paths as in the original method.
Fine-tuning improves temporal stability (please see videos), and sometimes improves visual quality (e.g., sharper results for \textsc{Spaceship}).
 Finally during inference we provide an 8 samples-per-pixel (spp) input image along with the all the buffers required (albedo, depth, etc.).

Our method demonstrates better temporal stability especially in parts of the scene where the noise in the input is higher such as the reflections in the \textsc{Spaceship} and \textsc{Veach Egg} scenes. ANF manages to successfully reconstruct parts of the scene where there is a big correlation between the input buffers and the final color, such as diffuse walls in \textsc{Living Room}, and parts where the light effect exists in the noisy input, highlights on the floor in \textsc{Living Room}. The limitation of ANF is clear in cases of complex light effects that do not appear in the noisy input due to the low spp and where the input buffers do not help, such as the red caustic in \textsc{Spaceship}, the bottle caustic in \textsc{Living Room} and the complex shadow of the glass egg \textsc{Veach Egg}. 

These effects have significant impact on the observed realism of the scene. However, they are completely missing from the path-traced+denoising solution, despite these effects being present in the ground truth images used for fine-tuning (provided in supplemental). \NEWTEXT{Quantitative results are shown in Tab.~\ref{tab:isik-table}; our method outperforms~Işık et. al.~\shortcite{isik2021anf} in \textsc{Spaceship} and \textsc{Living Room}. For \textsc{Veach Egg}, two metrics give our method lower score, even though we clearly capture indirect effects that are completely missing in Işık et. al.}

\begin{table}[!h]
	\includegraphics[width=\linewidth]{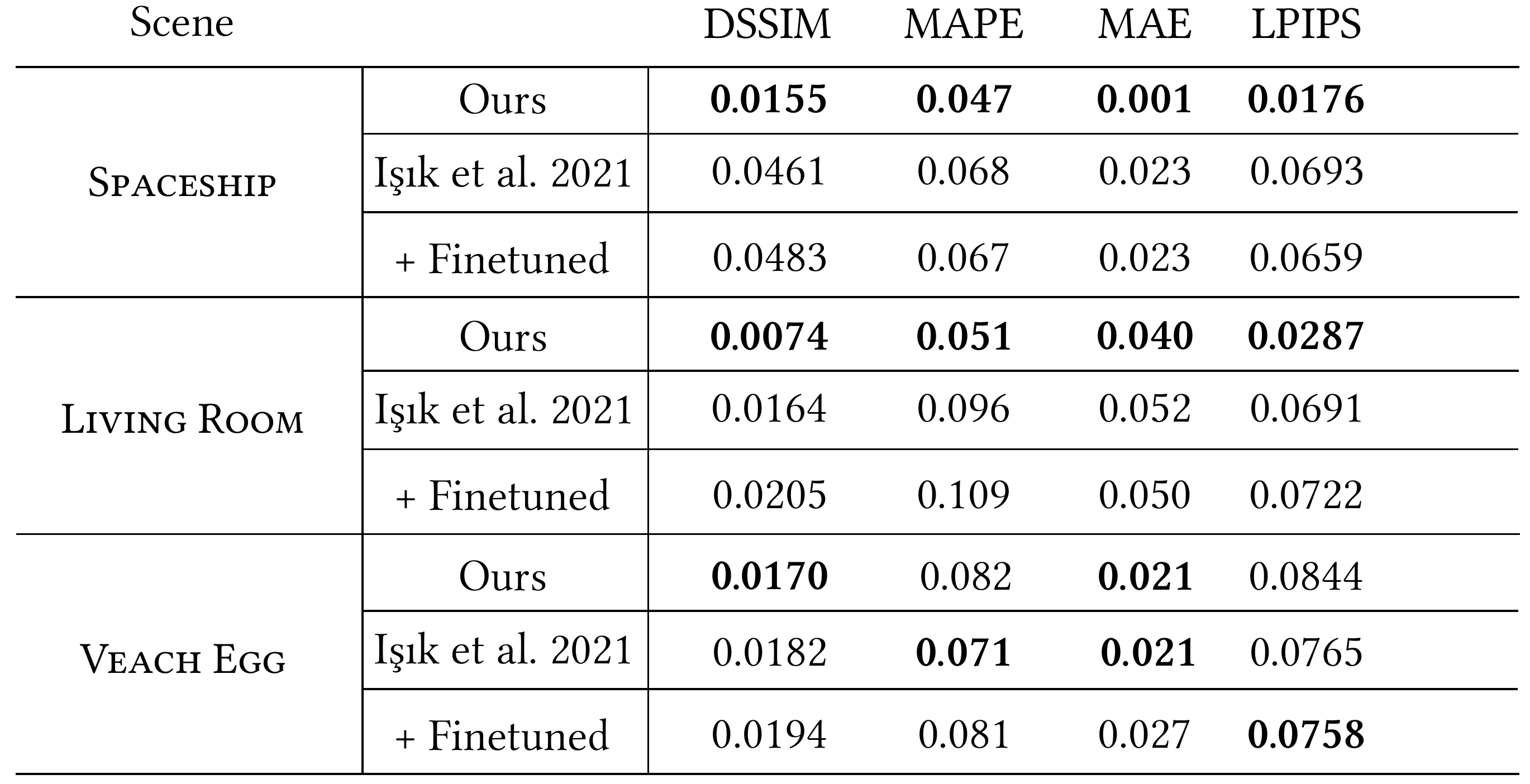}
	\caption{\label{tab:isik-table} \NEWTEXT{Quantitative results using 4 metrics for the configuration shown in Figure~\ref{fig:comp-denoiser}.}}
\end{table}

This illustrates one of the major strengths of our approach: the only way to render such hard light transport in a path-tracing context is to dramatically increase the number of samples per pixel.
In contrast our method encodes light transport in the neural network and uses the explicit scene representation vector to get information about such effects, such as the position of the glass egg or the cockpit light. As a result, we achieve interactive rendering with all effects present for the same training time.

\subsection{Evaluation}
\label{sec:evaluation}

We first study the effect of the number of variable dimensions, then present other ablations concerning different design choices of our method.

\begin{figure}[!h]
	\includegraphics[width=\linewidth]{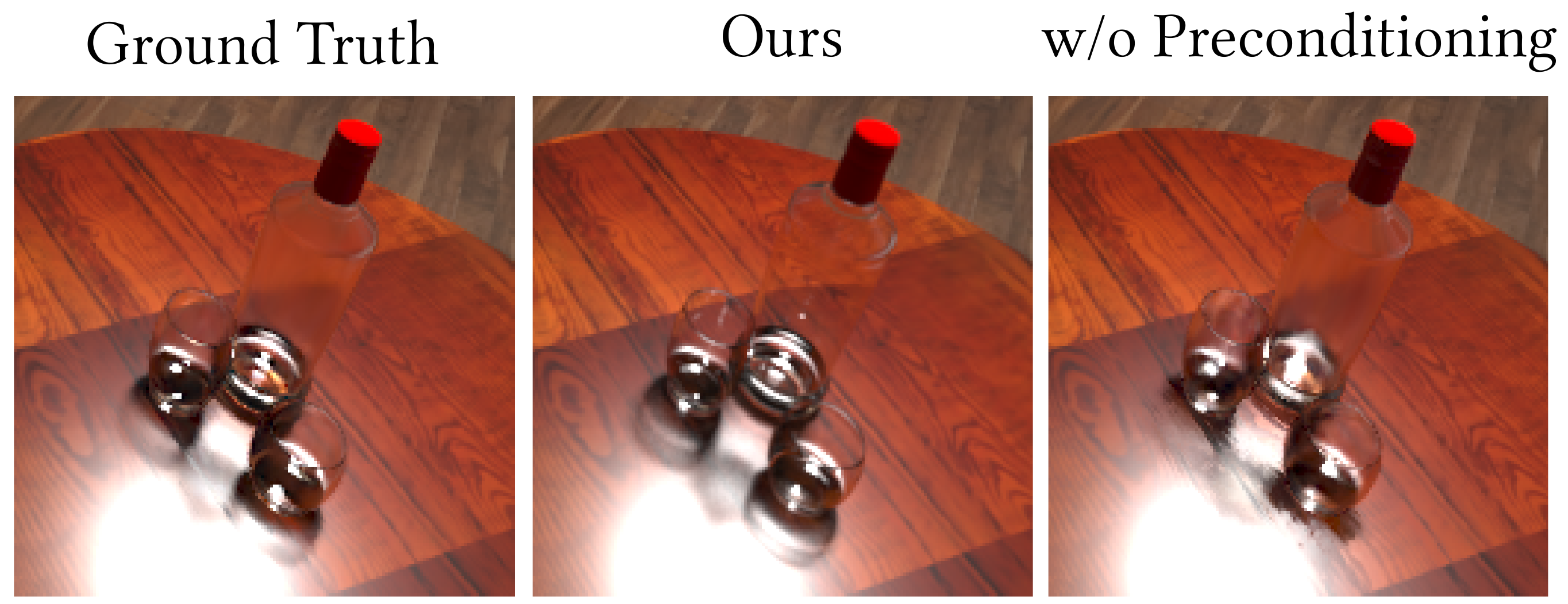}
	\caption{\label{fig:ablations-precondition}
		Position Preconditioning \NEWTEXT{allows the generator to ignore the high frequencies of the wood texture when it forms the shadows and caustic, resulting in better quality}.}
\end{figure}

\paragraph{Study of number of variable dimensions.}

We investigate the impact of the number of scene variables on our results. We trained our method and the uniform approach described above on 5 increasingly variable variants of the \textsc{Salon} scene. The first variant -- \textsc{Salon-5-dim} -- only varies viewpoint (5 dimensions), \textsc{Salon-7-dim} adds a movable set of furniture on the floor (7 dimensions). In \textsc{Salon-9-dim} the light source moves on the ceiling (9 dimensions). In \textsc{Salon-10-dim} the roughness of the wooden floor is also variable (10 dimensions). To demonstrate that some variables have a bigger impact on training time than others, e.g., light source position compared to changing albedo, we introduce \textsc{Salon-25-dim} which also varies the albedos of the furniture and walls (25 dimensions). In Fig.~ \ref{fig:variable-living} we see that while for \textsc{Salon-5-dim} the difference in validation loss between our method and the uniform sampling is small, \textsc{Salon-10-dim} demonstrates that the benefit of our method increases with higher numbers of variable dimensions. For this case, uniform search almost completely misses the important highlight on the glossy wooden floor due to the very specific configuration of parameters that create it. As a result Active Exploration is crucial for scenes with many variable elements, such as the ones used in production.

\paragraph{Ablation: preconditioning on position.}

In Fig.~\ref{fig:ablations-precondition} we show the difference in results between our full method and an ablation where position is concatenated with all other dimensions and fed directly to the network. We see clearly that the position preconditioning greatly improves overall performance. In the \textsc{Living Room} scene the albedo buffer for the \NEWTEXT{table} has high frequency variations due to the wood texture. Without the preconditioning it is hard for the PixelGenerator to learn to ignore this information when shading \NEWTEXT{the caustic and the shadow of the bottle}.
\NEWTEXT{Quantitative results in Tab.~\ref{tab:precondition} confirm this choice.}

\begin{table}[!h]
	\includegraphics[width=\linewidth]{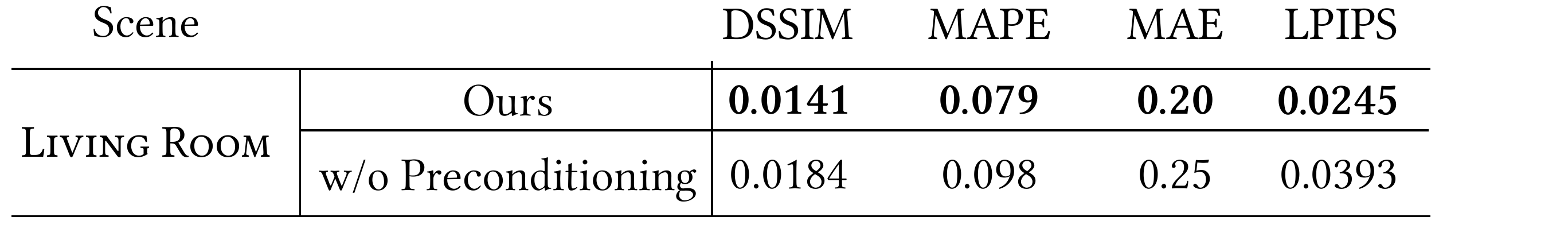}
	\caption{\label{tab:precondition} \NEWTEXT{Preconditioning improves the quantitative performance (see also Figure~\ref{fig:ablations-precondition}).}}
\end{table}

\paragraph{Ablation: increasing resolution.}

We next study the effect of progressively increasing resolution during training (Sec.~\ref{sec:resolution}).
In Fig.~\ref{fig:ablations-resolution}, we compare to an ablation where we do not increase resolution during training. We can see that the increase in resolution allows our active exploration to resolve high frequency effects such as reflections and shadows (lamp on the left) much more effectively.
\NEWTEXT{The corresponding quantitative results in Tab.~\ref{tab:ablation-resolution} confirm the improvement in quality.}

\begin{table}[!h]
	\includegraphics[width=\linewidth]{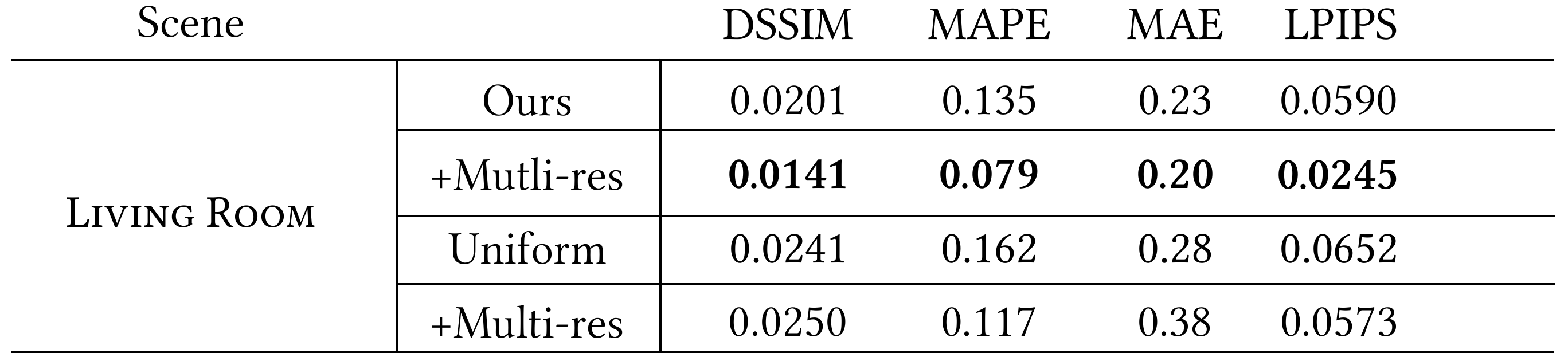}
	\caption{\label{tab:ablation-resolution} \NEWTEXT{Quantitative results illustrating the effect of resolution on error.}} 
\end{table}

\begin{figure}[!h]
\includegraphics[width=\linewidth]{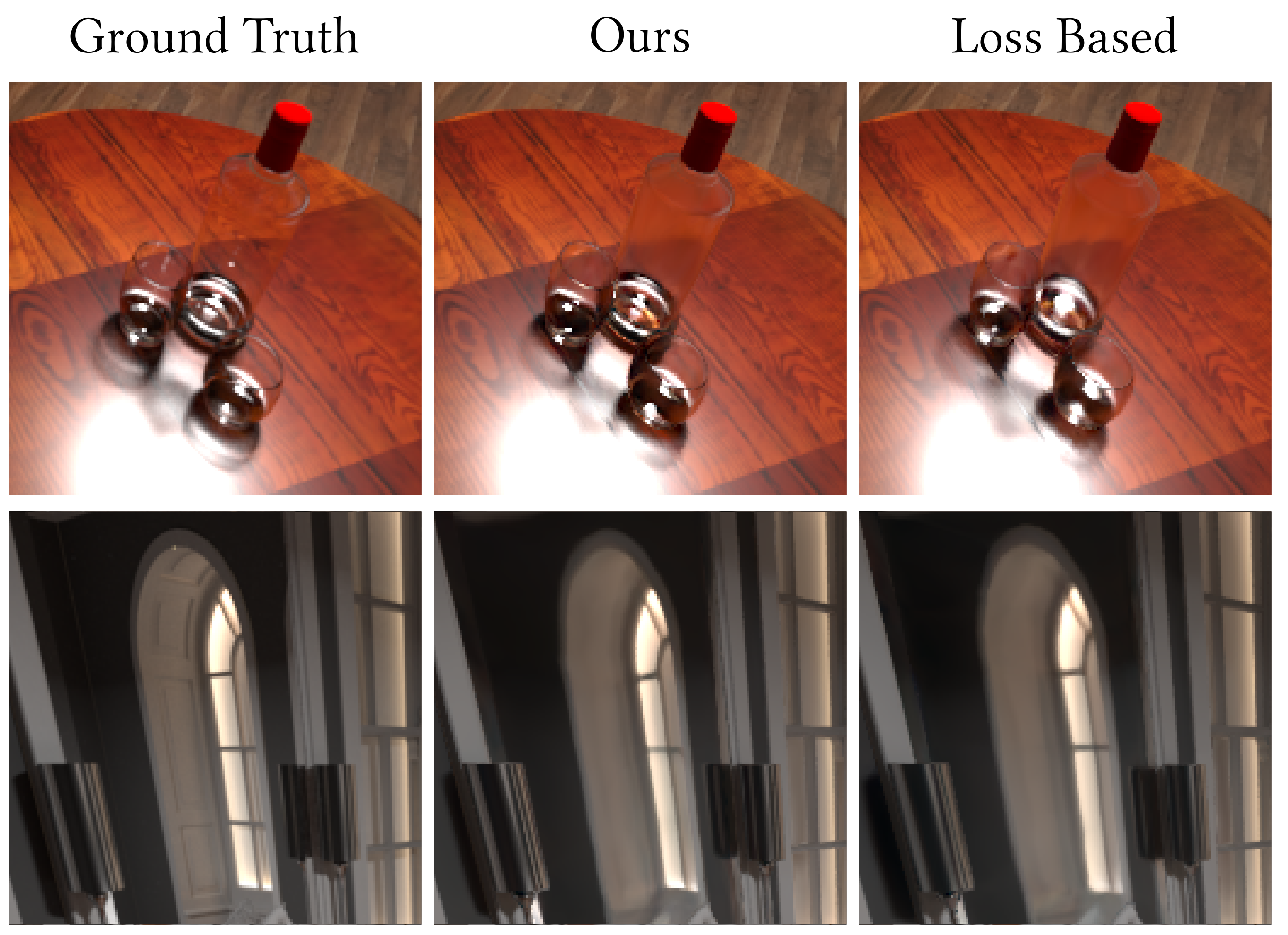}
		\caption{\label{fig:ablations-target}Use of the loss alone for the target function results in blurrier results.}
\end{figure}

\paragraph{Ablation: target function.}
In Fig. \ref{fig:ablations-target} \NEWTEXT{and Tab.~\ref{tab:target-ablation}}, we see that using only the loss for the target function degrades quality, since the training process gets stuck in local minima. \NEWTEXT{The MCMC finds configurations that cannot be improved anymore, such as the mirror reflection, and does not accept other states where the network could still improve, such as the bottle caustic. As a result the latter is lacking detail.}

\begin{table}[!h]
	\includegraphics[width=\linewidth]{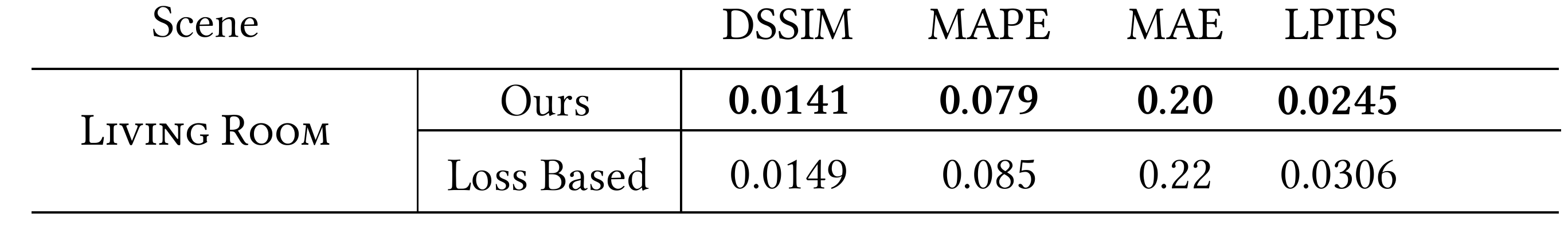}
	\caption{\label{tab:target-ablation} \NEWTEXT{Quantitative results using 4 metrics illustration the benefit of our choice of target function (see in Figure~\ref{fig:ablations-target}).}}
\end{table}

\section{Future Work, Limitations and Conclusion}

Despite providing interactive global illumination in dynamic scenes, our method is not without limitations; we discuss these below together with some avenues for future work before concluding.

\begin{figure}[!h]
	\includegraphics[width=\linewidth]{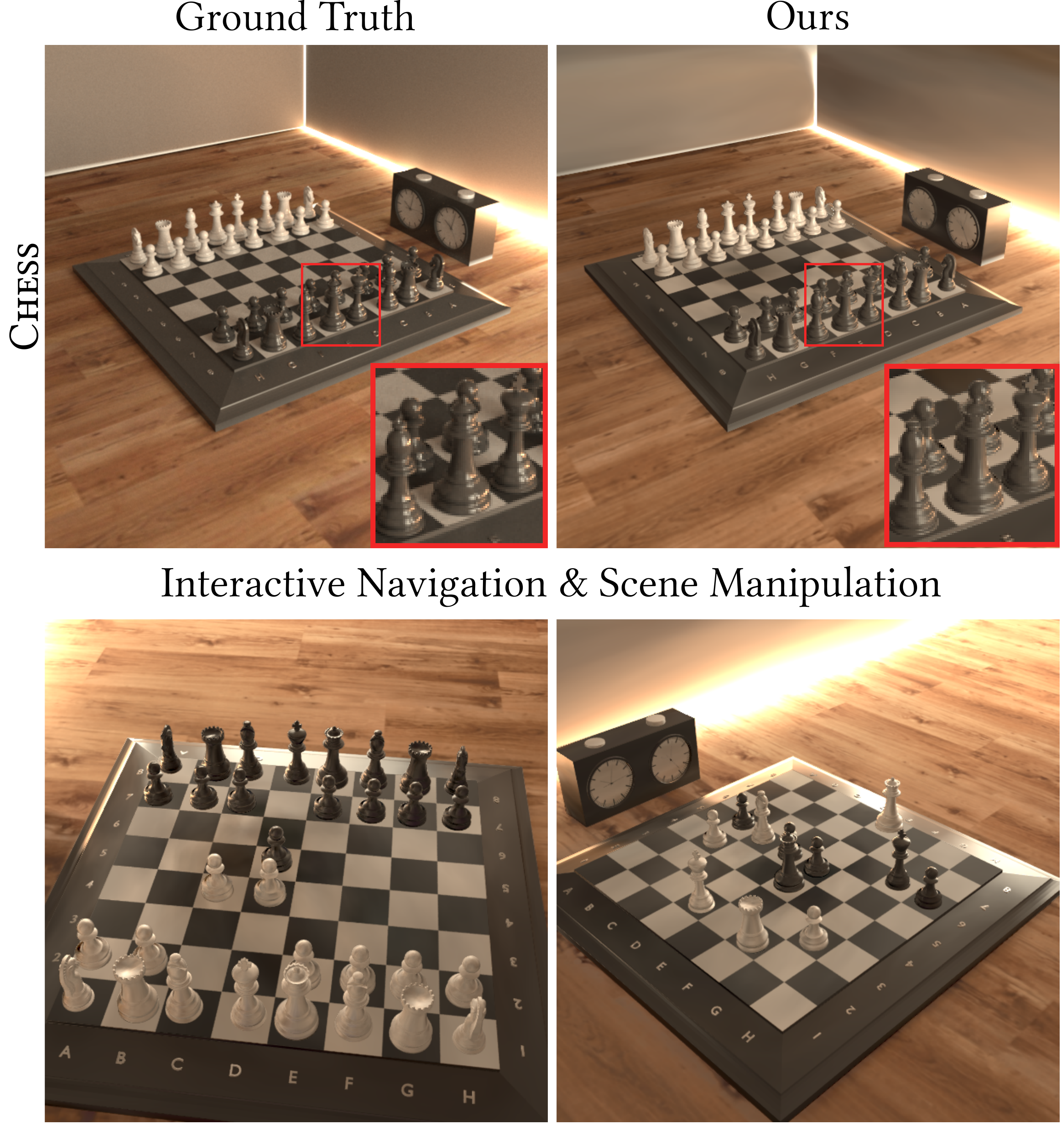}
	\caption{\label{fig:chess} The \textsc{Chess} scene tests the limits of our method with 128 variables. Each chess piece can be moved on the board, lifted and be captured. Our method still gives plausible results but is missing some shadows and highlights.}
\end{figure} 

Rendering using our unoptimized Python implementation currently runs at \OURFPS, including a 15ms overhead for generating G-buffers in Mitsuba -- which could be performed with hardware acceleration -- and an unoptimized inference step. We are confident that significant speedup can be achieved with further optimization. We chose to learn \emph{all} light paths, including mirror reflections. \NEWTEXT{While we achieve acceptable results in many cases, high-frequency effects may not be reproduced exactly. However, our approach can be used in a hybrid setting, using real-time ray-tracing for specular interactions as seen in Fig.~\ref{fig:ablation-specular}, overcoming this issue. If the use of path tracing is not an option, Neural Textures such as the ones used in \cite{thies2019deferred} could improve reflections in cases where the G-Buffers do not provide any meaningful information.}  

\NEWTEXT{Active Learning literature has explored many different metrics for deciding the value of each sample. In this work we explored two functions that can be efficiently computed in a single GPU training scenario but there are alternatives. In a multi GPU training scenario one option is to use a query by committee. Different copies of the model could be trained in parallel in each GPU and whenever a large step is performed all the models could be evaluated on the proposed state. Using the prediction variance of all the models' answers can be a good fit for a target function as it shows there is uncertainty on what the result should be. Additionally Bayesian Neural Networks with explicit access to uncertainty metrics could possibly be an option for our Active Exploration in the future.}

\NEWTEXT{One aspect we would like to explore in future work is how to take into account the importance/difficulty of each scene variable. From our tests different variables can have different impact on the scene's global illumination and can be harder/ easier to represent by the generator. In general, variables that create or control high frequencies, such as reflections and shadows, are much harder to learn than variables such as the color of emitters or objects. Explicit injection of this knowledge using some form of Importance Sampling could help reduce training times and improve quality. Another property of the variables that we do not handle explicitly is the difference in their ranges. Since we normalize each variable the network needs to learn to scale the normalized values accordingly to match their impact on the final rendering. For instance for two rotatable objects that can be rotated 360$^{\circ}$ and 15$^{\circ}$ respectively the network in both cases will receive values between 0 and 1 even though the first object will create much higher frequency shadows in that range. Finding a way to adapt the MCMC mutations to such range differences per variable could increase the efficiency of Active Exploration.}

The types of variables we demonstrated can have a big impact on the overall appearance of our scenes but they are simple to represent with a few floats (rotation, translation, roughness etc). In future work we would like to expand our method to variables that are difficult to represent such as the parametric deformations in \cite{sloan2005local}. We believe that finding  inventive ways to represent such variability (such as using the keyframe as a parameter) is a promising avenue of future research.

We still can require up to \MAXTRAIN~ of training time for a given scene, depending on the required quality and the number of variable parameters. As discussed earlier, the network architecture used can play a significant role in the quality of the results; it is possible that different architectures will further improve quality and thus training speed. Another possible extension could be to train with a set of variable parameters and allow fine-tuning of the network, e.g., allowing fast addition of a new object etc. Evidently, use of a faster path-tracer could also accelerate training. 

In future work, we believe our Active Exploration approach has significant promise for any neural rendering method (e.g.,~\cite{baatz21nerftex}) that trains on synthetic data, allowing potentially significant reduction in training time and improvements in quality.

\NEWTEXT{One limitation by design for our method is that we cannot handle thousands of variables. The scene representation vector is repeated to match the size of the G-Buffers so that the generator, which operates on a per pixel basis, is aware of the global state of the scene. For example, given 5000 variables (such as a variable texture) we would need to create a tensor of size 128x128x5000 that would be unmanageable in terms of memory. In such cases there is a need to encode this information in a different way, possibly through an encoder neural network. Our method, as shown in Figure~\ref{fig:chess}, can work with 128 variables with similar training times (18 hours) but the quality is lower than in simpler scenes (some missing highlights and shadows).}

In conclusion, we introduced a resolution-aware Active Exploration method that guides the sampling of the training data space, and a self-tuning sample reuse method that enables interleaved on-the-fly data generation and training. 

Our neural renderer, combined with our explicit scene instance parameterization vector, uses these contributions to capture hard light transport effects, allowing interactive exploration with full global illumination, including all light paths.

Using these elements we can render variable scenes after \TRAINTIMERANGE~of training, depending on scene and variation complexity and the quality required, including indirect lighting, shadows, transmission, glossy effects etc. Looking forward, we believe that our main contributions can be used beyond the precomputation scenario presented here: Active Exploration and self-tuning reuse could be used for future solutions that can provide data online, e.g., with real-time path tracing.

\begin{acks}
	This research was funded by \grantsponsor{788065}{ERC Advanced grant FUNGRAPH No 788065 (\url{http://fungraph.inria.fr})}{http://fungraph.inria.fr}. The
	authors are grateful to the OPAL infrastructure from Université
	Côte d’Azur for providing resources and support.
	The authors would also like to thank the anonymous referees for
	their valuable comments and helpful suggestions, as well as Georgios Kopanas and Felix H{\"a}hnlein for fruitful discussions. 
\end{acks}

\bibliographystyle{ACM-Reference-Format}
\bibliography{neural-rendering-paper}
	
\end{document}